\documentclass[fleqn,usenatbib]{mnras}

\usepackage{newtxtext,newtxmath}

\usepackage[T1]{fontenc}

\DeclareRobustCommand{\VAN}[3]{#2}
\let\VANthebibliography\thebibliography
\def\thebibliography{\DeclareRobustCommand{\VAN}[3]{##3}\VANthebibliography}

\usepackage{graphicx}	
\usepackage{amsmath}	

\usepackage{amssymb}	

\usepackage{pdflscape}
\usepackage{newtxtext,newtxmath}
\usepackage{graphicx}
\usepackage{siunitx}
\usepackage{mathtools}
\usepackage{mathrsfs}
\usepackage{physics}
\usepackage{ulem}
\usepackage{soul}
\usepackage[dvipsnames]{xcolor} 

\title[Weakened magnetic braking through DR]{Transition of latitudinal differential rotation as a possible cause of weakened magnetic braking of solar-type stars}

\author[T.Tokuno et al.]{
Takato Tokuno,$^{1,2}$\thanks{E-mail: tokuno-takato@g.ecc.u-tokyo.ac.jp (TT)}
Takeru K. Suzuki,$^{1,2}$
and Munehito Shoda$^{3}$
\\
$^{1}$ Department of Astronomy, School of Science, The University of Tokyo, Bunkyo-ku, Tokyo 113-0033, Japan \\
$^{2}$ School of Arts \& Sciences, University of Tokyo, 3-8-1, Komaba, Meguro, 153-8902 Tokyo, Japan \\
$^{3}$ Department of Earth and Planetary Science, School of Science, The University of Tokyo, Bunkyo-ku, Tokyo 113-0033, Japan \\
}

\date{Accepted XXX. Received YYY; in original form ZZZ}

\pubyear{2022}

\begin{document}
\label{firstpage}
\pagerange{\pageref{firstpage}--\pageref{lastpage}}
\maketitle

\begin{abstract}
We investigate the role of latitudinal differential rotation (DR) in the spin evolution of solar-type stars. Recent asteroseismic observation detected the strong equator-fast DR in some solar-type stars. Numerical simulations show that the strong equator-fast DR is a typical feature of young fast-rotating stars and that this tendency is gradually reduced with stellar age. Incorporating these properties, we develop a model for the long-term evolution of stellar rotation. The magnetic braking is assumed to be regulated dominantly by the rotation rate in the low-latitude region. Therefore, in our model, stars with the equator-fast DR spin down more efficiently than those with the rigid-body rotation. We calculate the evolution of stellar rotation in ranges of stellar mass, $0.9 \, \mathrm{M}_{\odot} \le M \le 1.2\, \mathrm{M}_{\odot}$, and metallicity, $0.5\, \mathrm{Z}_{\odot} \le Z \le 2\, \mathrm{Z}_{\odot}$, where $\mathrm{M}_{\odot}$ and $\mathrm{Z}_{\odot}$ are the solar mass and metallicity, respectively. Our model, using the observed torque in the present solar wind, nicely explains both the current solar rotation and the average trend of the rotation of solar-type stars, including the dependence on metallicity. In addition, our model naturally reproduces the observed trend of the weakened magnetic braking in old slowly rotating solar-type stars because strong equator-fast DR becomes reduced. Our results indicate that latitudinal DR and its transition are essential factors that control the stellar spin down.
\end{abstract}




\section{Introduction}

\label{sec:intro}

Main-sequence stars in the lower-mass side possess a convective envelope, which amplifies magnetic field and induces a variety of magnetic activity \citep[e.g.][]{Strugarek2017Sci,Hotta_Iijima2020MNRAS,Fan2021LRSP}. Observations of such `solar-type' stars with various ages enable us to track the expected long-time evolution of our Sun \citep[][]{Ribas2005ApJ,Wood2014ApJ,Vidotto2014MNRAS,Brun2017LRSP,OFionnagain2019MNRAS,Toriumi2022ApJS}. 
In recent years, the stellar magnetic activity has also attracted great attention in terms of the impact on the environments of planets \citep[][]{Shibata2013PASJ,Yamashiki2019ApJ,Viddoto2021LRSP, Temmer2021LRSP}.

Magnetised stellar winds emanate from a high-temperature corona of solar-type stars \citep[e.g.][]{Parker1958ApJ,Kopp1976SoPh,Velli1994ApJ,Cranmer&Saar2011ApJ,Suzuki2013PASJ}. They carry off not only the mass but also the angular momentum (AM hereafter) of a star; solar-type stars generally spin down through this magnetic braking mechanism \citep{Weber1967ApJ,Kawaler1988ApJ,Matt2012ApJ}. 
The spin-down of solar-type stars can be utilised to estimate the stellar age via gyrochronology \citep[][]{Barnes2003ApJ,Garcia2014AandA}. The stellar spin is also believed to govern the magnetic activity through the dynamo mechanism \citep[e.g.][]{Parker1955ApJ,Hotta2015ApJ,Brun2022ApJ} and affect rotationally induced chemical mixing in the stellar interior \citep[e.g.][]{Zahn1992A&A,Palacios2003A&A}. 
\citet{Skumanich1972ApJ}
discovered a simple relation, $\Omega  \propto t^{-1/2}$, between the angular velocity, $\Omega$, and the age, $t$, of observed G-type stars. The overall trend of the Skumanich's law is also confirmed by later observations \citep[][]{Barnes2007ApJ,Angus2015MNRAS}.

Recent space telescope missions , CoRoT \citep{Baglin:2006aa} and \textit{Kepler} \citep{Borucki:2010aa}, aiming primarily at transit observations of exoplanets, have also been greatly contributing to our understandings of magnetic and rotational properties of solar-type stars \citep[e.g.][]{Maehara2012Nature,Notsu2013ApJ,Namekata2019ApJ}. The rotational period can be obtained from the periodic variation of a light curve that is considered to mainly arise from dark star spots appearing into and out of view by rotation \citep[][]{McQuillan2014ApJS,Namekata2020ApJ,Masuda2022ApJ}. In addition, high-quality light curves enable us to perform asteroseismic analyses by using oscillations detected at the stellar surface. An advantage of solar-type stars is that we can use a large number of $p$-modes, which correspond to acoustic oscillations, excited in a surface convection zone \citep[e.g.][]{Unno1989,Aerts2010,Garcia2019LRSP}. 
The basic stellar parameters, such as mass and age, of solar-type stars can be determined by the asteroseismic modelling \citep[e.g.][]{SilvaAgirre2017ApJ,AgBoKo2022MNRAS}.

The time evolution of the stellar AM, $J$, is described by 
\begin{align}
    \frac{\mathrm{d}J}{ \mathrm{d}t}=-\tau_\mathrm{w}, 
    \label{eq:AMev_intro}
\end{align}
where $\tau_\mathrm{w}$ denotes the wind torque. The key here is how to determine $\tau_\mathrm{w}$ as a function of basic stellar parameters. A simple prescription is to assume the power-law dependence on rotation, $\tau_\mathrm{w} \propto \Omega^{p+1}$, with a constant scaling index, $p+1$, which gives $\Omega \propto t^{-1/p}$. We note that the Skumanich's law \citep{Skumanich1972ApJ} is recovered when $p=2$. The scaling index is inferred from observations of the rotation periods of stars with different ages \citep[][]{Skumanich1972ApJ,Johnstone2021A&A} 
and numerical simulations \citep[][]{Reville2016ApJ,Shoda2020ApJ}. For example, previous models of the spin-down that adopt the observational scaling index explain the basic observed trend of solar-type stars that are younger than the Sun \citep[e.g.][]{Matt2015ApJ}. However, there are still two major unsolved problems remained.

First, the empirical torque formulation that reproduces the observed spin-down of both the Sun and stars yields a solar-wind torque \citep[$\tau_\mathrm{w,\odot,M15} = 6.3 \times 10^{30} \, \mathrm{erg}$;][]{Matt2015ApJ} twice as large as the direct measurement by the WIND spacecraft \citep[$\tau_\mathrm{w,\odot,F19} = 3.3 \times 10^{30} \, \mathrm{erg}$;][]{Finley2019ApJ_Direct}. 
The WIND observation is consistent with the recent observation by Parker Solar Probe \citep[][]{Fox2016SSRv}, which gives $\tau_\mathrm{w,\odot,F20}=(2.6-4.2)\times 10^{30} \, \mathrm{erg}$ \citep{Finley2020ApJ}. 
We note that one dimensional magnetohydrodynamical (1D MHD) simulations in the framework of Alfv\'en-wave-driven stellar winds support the observed value \citep[][]{Shoda2020ApJ}. It is essentially important to solve this discrepancy and construct a self-consistent model for the wind torque.

Second, stars older than the Sun spin down considerably slower than the conventional spin-down law \citep{vanSaders2016Nature, Hall2021NatAs, Masuda2022MNRAS_Inferring}.
The breakdown of the spin-down law means that a regime change in the magnetic braking process occurs around the solar age.
The weakened magnetic braking is possibly caused by the change in the magnetic geometry from a dipole field by large-scale dynamo to higher-order fields by small-scale dynamo \citep{vanSaders2016Nature,Metcalfe2022ApJ}.
However, there might be other mechanisms at play.

In this paper, we consider an effect of differential rotation (DR), especially latitudinal differential rotation (LDR),  as a possible mechanism to resolve these two problems. Theoretical models for the long-term evolution of stellar AM often assume the one-zone \citep[e.g.][]{Matt2015ApJ} or two-zone \citep{MacGregor1991ApJ} solid-body rotation, where the two zones consist of the radiative core and the convective envelope. The assumption of the solid-body rotation is valid only if the stellar LDR is weak enough. Therefore, it is particularly important to take into account the stellar LDR for fast rotating stars.

The asteroseismic method can be utilised to detect the stellar LDR through the rotational splitting of frequency modes \citep{Gizon2004SoPh}. Using \textit{Kepler} data, \citet{Benomar2018Sci} detected the LDR profiles for 13 solar-type stars out of 40 samples. Their remarkable finding is that these stars show the strong equator-fast DR. 3D MHD simulations \citep[][]{Kapyla2011A&A,Gastine2014MNRAS,Brun2022ApJ} show that the stellar LDR is divided into two types: fast rotators ($\mathrm{Ro} \lesssim \mathrm{Ro}_\odot$) take equator-fast DR, while slow rotators ($\mathrm{Ro} \gtrsim \mathrm{Ro}_\odot$) have pole-fast DR, where $\mathrm{Ro}$ is the Rossby number defined as the ratio of the rotation period to the convective overturn timescale and $\mathrm{Ro}_\odot$ is the corresponding solar value. We note that the transition from equator-fast DR to pole-fast DR appears to occur at $\mathrm{Ro} \sim \mathrm{Ro}_\odot$, which is roughly consistent with the asteroseismic observations at least in a qualitative sense \citep[cf.][]{Saar2011IAUS,Benomar2018Sci}.

Moreover, \citet{Metcalfe&vanSaders2017SoPh} show the transition almost matches the declining trend of the stellar magnetic activity, which implies that the Sun is now under the transition phase. Hence, we should be careful when constructing theoretical models for the long-term evolution of stellar AM because a simple extrapolation of the solar properties to a wide range of stellar rotation would not be a plausible assumption. Recently, \citet{Ireland2022ApJ} also show that the weak LDR that is comparable to the solar DR still affects the quantitative evaluation of the integrated wind torque by using the analytical solution of \citet{Weber1967ApJ} and their own numerical simulations. A qualitatively similar result is also obtained in a recent work by \citet{FinleyBrun2023arXiv}. In summary, it is worth pursuing the effect of LDR in the spin-down of solar-type stars.

On the basis of these results, in this paper, we construct a phenomenological model on the AM evolution of solar-type stars explicitly considering LDR. Then, we investigate whether LDR can be a key to solve the above-mentioned two problems on the stellar spin down. The novelty of our model calculations is that we can directly derive the relation between the stellar spin-down and LDR.

The outline of this paper is as follows. In Section \ref{sec:Model}, we first propose our model for the wind torque that explicitly takes into account LDR. Utilising this model, we formulate an analytic model for the long-term evolution of stellar AM. After the brief introduction of observational data (Section \ref{sec:observation}), we show main results of our model calculation, through comparison to observations (Section \ref{sec:Result}). Discussions and conclusions are presented in Sections \ref{sec:Discussion} and \ref{sec:Conclusion}, respectively.

\section{Model}
\label{sec:Model}

In this section, we describe our model that incorporates the effect of LDR in the evolution of stellar AM. We first introduce the physical picture of the model, particularly focusing on the connection between the LDR and the removal and internal transport of AM (Section \ref{subsec:picture}). In Section \ref{subsec:DR_model}, we formulate analytical expressions of the LDR that explain the asteroseismic observation by \citet{Benomar2018Sci} and the numerical simulation by \citet{Brun2022ApJ}. In Section \ref{subsec:Wind_model}, we construct a model for the wind torque by extending the formulation developed by \citet{Matt2015ApJ}. Finally, we describe our model and specific procedures for the long-term evolution of stellar AM in Section \ref{subsec:AM_evo}.

\subsection{Overall picture of LDR effect} \label{subsec:picture}

\begin{figure}
    \includegraphics[width=\columnwidth]{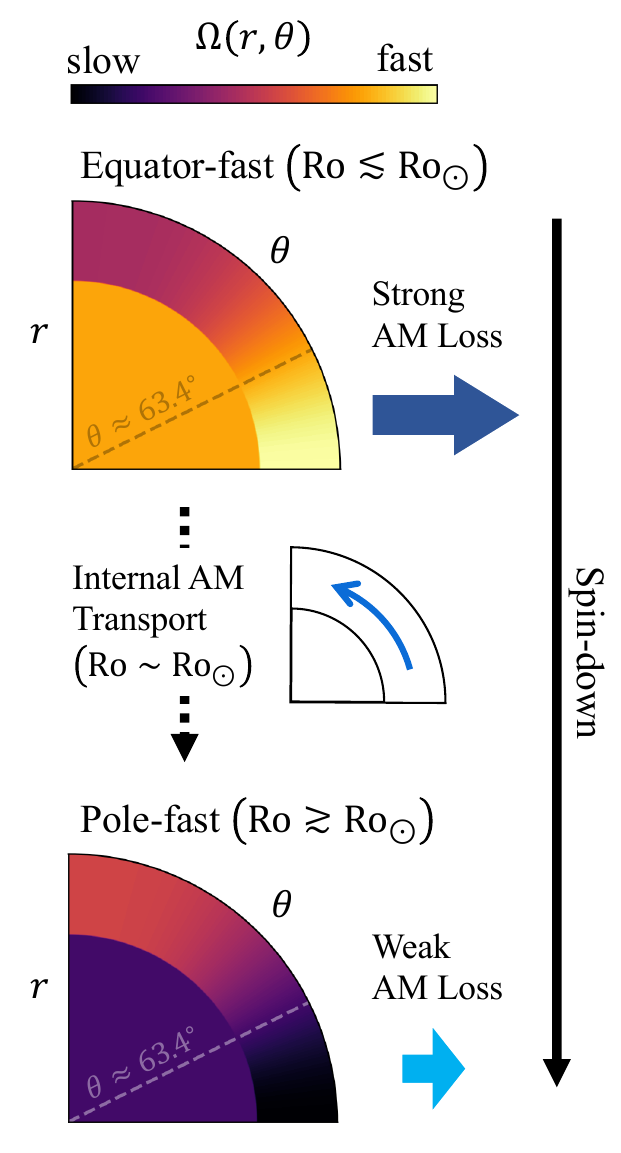}
    \caption{The schematic picture of the LDR effect on the stellar spin-down. The top horizontal colour-bar indicates the stellar angular velocity, $\Omega(r,\theta)$ (the analytical model is given by equation \ref{eq:DRprofile_model}), where $r$ and $\theta$ denote the distance from the centre and co-latitude, respectively. The upper and lower quarter sector represents a star with equator-fast DR (when $\mathrm{Ro} \lesssim \mathrm{Ro}_\odot$) and pole-fast DR (when $\mathrm{Ro} \gtrsim \mathrm{Ro}_\odot$), respectively. The middle quarter circle indicates a star in the transition from equator-fast DR to pole-fast DR at $\mathrm{Ro} \sim \mathrm{Ro}_\odot$ ; the light-blue arrow illustrates the internal transport of AM.  Under the assumption that the magnetic braking is regulated by the low-latitude rotation speed, AM is removed more strongly (weakly) under the equator-fast DR (pole-fast DR). The inclined dashed line corresponds to  the co-latitude where the angular velocity of the envelope is equal to that of the core, which is $\theta = \arccos(1/\sqrt{5}) \approx 63.4^{\circ}$ (see Section \ref{subsec:DR_model}).}
    \label{fig:Schematic_model}
\end{figure}

Fig.~\ref{fig:Schematic_model} represents a schematic picture of the LDR effect on the stellar spin-down. 
We use the Rossby number as the control parameter of our model for the stellar rotation, following \citet{Noyes1984ApJ} and \citet{Stepien1994A&A}. Here, we adopt an observationally determined empirical Rossby number, 
\begin{equation}
  \mathrm{Ro} \equiv  \frac{2\pi}{\Omega_{\rm ave} \tau^{\rm CS}_{\rm cz}}, \label{eq:def_rossby}
\end{equation}
where $\Omega_{\rm ave}$ represents the averaged angular velocity of the star and  $\tau^{\rm CS}_{\rm cz}$ denotes the convective turn-over time, which is estimated from the effective temperature, $T_{\rm eff}$, as \citep{Cranmer&Saar2011ApJ}
\begin{equation}
  \tau^{\rm CS}_{\rm cz} \equiv 314.24 \times \exp\left[-\frac{T_{\rm eff}}{1952.5 \, {\rm K}} - \left( \frac{T_{\rm eff}}{6250 \, {\rm K}} \right)^{18} \right]+0.002 \, {\rm d}. \label{eq:timescale_overturn}
\end{equation}
Note that equation (\ref{eq:def_rossby}) well explains Ro obtained by numerical simulations at least in a proportional sense \citep{Brun2022ApJ}.

Solar-type stars initially rotate rapidly ($\mathrm{Ro} \ll \mathrm{Ro}_{\odot}$) at the zero-age main sequence and spend their early life with equator-fast LDR \citep[][]{Benomar2018Sci}. In addition to the removal of stellar AM by the magnetised stellar winds, the meridional transport of the AM from the equatorial region to the polar regions in the stellar interior \citep[][]{Amard2016A&A,Hanasoge2022LRSP} slows down the rotation velocity in the low-latitude region without net AM loss of the star. As a result, the LDR changes from equator-fast DR to pole-fast DR as the stars spin down when $\mathrm{Ro} \gtrsim \mathrm{Ro}_\odot$ \citep[][]{Brun2022ApJ}. We incorporate these evolutionary trends of the stellar AM and LDR in our model (Sections \ref{subsec:DR_model} and \ref{subsec:Wind_model}).

The key ingredient in our prescription is the effect of LDR in the wind torque. Specifically, we focus on the situation where the wind torque is mainly determined by the rotation rate in the low-latitude region. This is because the stellar wind from there is supposed to carry off AM more effectively from naive geometrical considerations \citep[e.g.][]{Weber1967ApJ}. When the wind torque is positively dependent on the equatorial angular velocity, the outward transport of AM from a star with equator-fast DR (pole-fast DR) is stronger (weaker) than that with solid-body rotation. On the other hand, the quantitative properties of the wind torque are determined by the location of the footpoint of open magnetic field lines, which should be calculated from the detailed configuration of the stellar magnetic field that interacts with stellar winds \citep[][]{Reville2015ApJ,Garraffo2015ApJ,Finley2018ApJ_Geo2, FinleyBrun2023arXiv}. 
In order to consider this ambiguity, we introduce a parameter that expresses the dependence of the wind torque on LDR and test various possibilities
(Section \ref{subsec:Wind_model}).


\subsection{LDR model}  \label{subsec:DR_model}

We introduce how we treat LDR analytically. The standard spherical coordinate system $(r,\theta, \varphi)$ is used to describe the spatial distribution. For simplicity, we assume that the stellar internal structure is spherically symmetric in its equilibrium and that the stellar rotation is axisymmetric.

From the result of helioseismology, the solar angular velocity profile $\Omega_\odot (r,\theta)$ is obtained with high precision \citep[e.g.][]{Thompson2003ARA&A}. In contrast, the information obtained from asteroseismic observations of solar-type stars is quite limited, and thus, several assumptions are required to infer the profile of LDR.
Following \citet{Benomar2018Sci} and \citet{Bazot2019AandA}, we assume the two-zone LDR model. Letting $\Omega_\mathrm{core}$ and $\Omega_\mathrm{env}$ be the angular velocity of the core and envelope of a star, respectively, the poloidal profile of the stellar angular velocity, $\Omega(r,\theta)$, obeys
\begin{gather}
\Omega(r,\theta) = 
\begin{cases}
\Omega_\mathrm{core} \ (\mathrm{Const.}) & \mathrm{for} \ r \leq R_\mathrm{core} \\ 
\Omega_\mathrm{env}(\theta) = \Omega_\mathrm{core} +\Delta \Omega_\mathrm{core} (1 - 5 \cos^2 \theta) & \mathrm{for} \ r > R_\mathrm{core}
\end{cases}.
\label{eq:DRprofile_model}
\end{gather}
Here, $R_\mathrm{core}$ denotes the boundary between the core and envelope and $\Delta \Omega_\mathrm{core}$ represents the strength of LDR defined as
\begin{gather}
    \Delta \Omega_\mathrm{core} \equiv \Omega_\mathrm{eq} - \Omega_\mathrm{core}, \label{eq:def_Delta_Omega}
\end{gather}
where $\Omega_\mathrm{eq} \equiv \Omega_\mathrm{env}(\pi/2)$ is the angular velocity at the equator
    \footnote{Note on $\Delta \Omega$ in previous works: \citet{Benomar2018Sci} and \citet{Brun2022ApJ} adopt $\Delta \Omega_{45^\circ} \equiv \Omega_\mathrm{eq} - \Omega_\mathrm{env}(\pi/4)$ and $\Delta \Omega_{30^\circ} \equiv \Omega_\mathrm{eq} - \Omega_\mathrm{env}(\pi/3)$.}
. In this model, $\Omega_\mathrm{env}(\theta)$ equals $\Omega_\mathrm{core}$ at $\theta_\mathrm{core} = \arccos (1/\sqrt{5}) \approx 63.4^{\circ}$ (Fig.~\ref{fig:Schematic_model}), which is roughly consistent with the observed value in the solar case \citep[][]{Thompson2003ARA&A}. In this paper, we adopt the solar values, $\Omega_\mathrm{core}= 436.5 \, \mathrm{nHz / 2\pi}$ and $\Delta \Omega_\mathrm{core} = 15.0 \, \mathrm{nHz / 2\pi} $, which are obtained from fitting $\Omega_\mathrm{env}$ to the observed data in $\theta_\mathrm{core} \leq \theta \leq  90^\circ$ by \citet{Ulrich1988SoPh}.

It should be noted that equation (\ref{eq:DRprofile_model}) guarantees the simple relation, 
\begin{gather}
    J = I \Omega_\mathrm{core},  \label{eq:def_AM}
\end{gather} 
between the total AM and the moment of inertia, $I$, of a star. 
We therefore simply assume $\Omega_{\rm ave} = \Omega_{\rm core}$ when calculating the Rossby number via equation (\ref{eq:def_rossby}). 
In deriving equation (\ref{eq:DRprofile_model}), we assume that 1. the core is rigidly rotating and 2. the terms higher than the second order with respect to $\cos \theta$ are negligibly small. The assumption 1. is found to be satisfied in the Sun \citep[][]{Schou1998ApJ,Thompson2003ARA&A,Eff-Darwich2013SoPh} and solar-type stars \citep[][]{Benomar2015MNRAS, Saio2015MNRAS}. 
The assumption 2. possibly yields a significant deviation between the model and the observed solar profile, in particular in the high-latitude regions. However, the error is relatively small in the low-latitude region where we focus on. $\Delta \Omega_\mathrm{core}$ is, indeed, defined as the difference between the angular velocities at the low-latitude, where $\theta = 90^\circ$ and  $\theta = \theta_\mathrm{core} \approx 63.4^\circ$ (see equation \ref{eq:DRprofile_model}).

$\Delta \Omega_\mathrm{core}$ is given as a function of Ro (equation~\ref{eq:def_rossby}) to reproduce observations \citep{Saar2011IAUS,Benomar2018Sci,Bazot2019AandA} and simulations \citep{Brun2022ApJ}.
We divide the stellar rotation into three regimes: `fast rotators' ($\mathrm{Ro}_\mathrm{crit} \gtrsim \mathrm{Ro}$), `moderate rotators' ($\mathrm{Ro}_\mathrm{tran} \gtrsim \mathrm{Ro} \gtrsim \mathrm{Ro}_\mathrm{crit}$ ), and `slow rotators' ($\mathrm{Ro} \gtrsim \mathrm{Ro}_\mathrm{tran}$), where $\mathrm{Ro}_\mathrm{crit}$ and $\mathrm{Ro}_\mathrm{tran}$ denote the Rossby numbers at the transitions between saturated and unsaturated regimes and between equator-fast DR and the pole-fast DR, respectively. In each regime, $\Delta \Omega_\mathrm{core}$ is prescribed as follows.
\begin{gather}
    \frac{\Delta \Omega_\mathrm{core} }{\Omega_\mathrm{eq}} = 
    \begin{cases}
        \alpha_\mathrm{ef} (\mathrm{Ro}/\mathrm{Ro}_\mathrm{crit})^{2}   & \text{for $\mathrm{Ro}_\mathrm{crit} \gtrsim \mathrm{Ro}$ (Fast)}, \\
        \alpha_\mathrm{ef} \ (\mathrm{Const.}>0) & \text{for $\mathrm{Ro}_\mathrm{tran} \gtrsim \mathrm{Ro} \gtrsim \mathrm{Ro}_\mathrm{crit}$ (Moderate),} \\
        \alpha_\mathrm{pf} \ (\mathrm{Const.}<0) & \text{for $\mathrm{Ro} \gtrsim \mathrm{Ro}_\mathrm{tran}$ (Slow),}
    \end{cases}
    \label{eq:Omega_Rossby_trend}
\end{gather}
where $\alpha_\mathrm{ef}$ and $\alpha_\mathrm{pf}$ as the maximum magnitudes of equator-fast DR and pole-fast DR, respectively. For the purpose of numerical integration, the three regimes are connected smoothly by
\begin{gather}
    \frac{\Delta \Omega_\mathrm{core}}{\Omega_\mathrm{eq}}=\left[\frac{\alpha_{\mathrm{ef}}\left(\mathrm{Ro} / \mathrm{Ro}_{\mathrm{tran}}\right)^{-n}}{1+\left(\mathrm{Ro} / \mathrm{Ro}_{\text {tran }}\right)^{-n}}+\frac{\alpha_{\mathrm{pf}}}{1+\left(\mathrm{Ro}/ \mathrm{Ro}_{\text {tran }}\right)^{-n}}\right] \notag \\
    \hspace{60pt} \times \left(\frac{1}{1+\left(\mathrm{Ro} / \mathrm{Ro}_{\mathrm{crit}}\right)^{- m}}\right)^{2/m} , \label{eq:Omega_Rossby_formula}
\end{gather}
where we set smoothing indices, $n=m=20$, throughout the paper. The critical Rossby number is set to $\mathrm{Ro}_\mathrm{crit} = \mathrm{Ro}_\odot / \chi$ for $\chi =10$ \citep[see][]{Matt2015ApJ,See2019ApJ}.  $\mathrm{Ro}_\mathrm{tran}$ is determined to reproduce $\Delta \Omega_\mathrm{core} / \Omega_\mathrm{eq} = \Delta \Omega_\mathrm{core,\odot} / \Omega_\mathrm{eq,\odot}$ at $\mathrm{Ro} = \mathrm{Ro}_\odot$, because the Sun is considered to be in the transition phase.
To date, no stars with pole-fast DR have yet been detected (discussed in Section \ref{sec:Discussion}). Hence, referring to the simulation data by \citet{Brun2022ApJ}, we choose $\alpha_\mathrm{pf} = -3 \Delta \Omega_\mathrm{core,\odot} / \Omega_\mathrm{eq,\odot}$.
We leave $\alpha_\mathrm{ef}$ as the only free parameter of equation (\ref{eq:Omega_Rossby_formula}) that determines the LDR from Ro.

\begin{figure}
    \centering
	\includegraphics[width=\columnwidth]{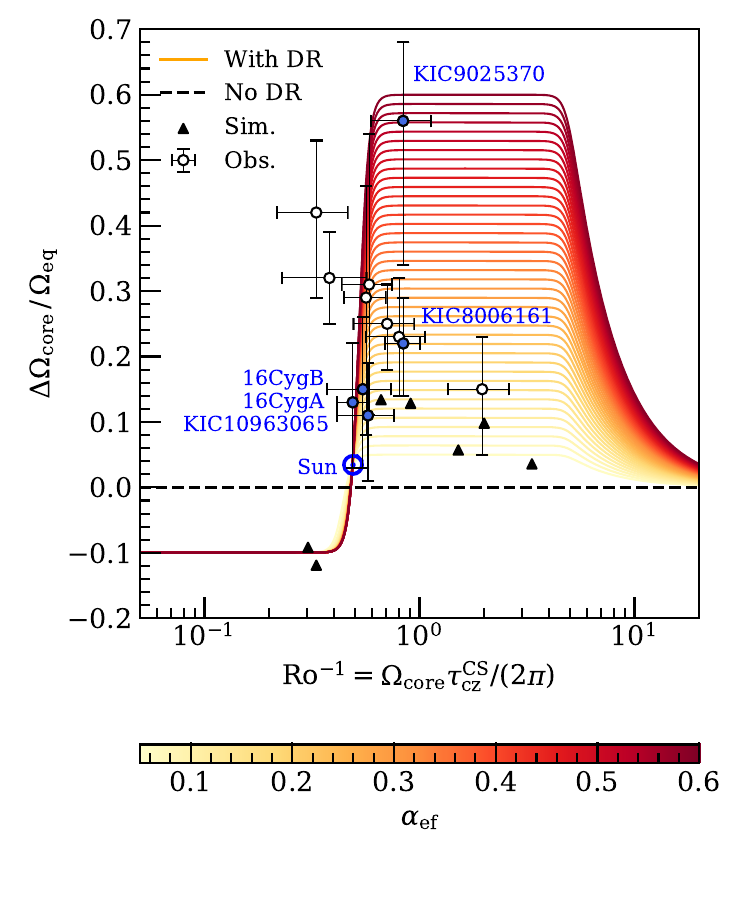}
    \caption{Comparison of the evolution of LDR (solid lines) with observations and numerical simulations. The abscissa  and ordinate represent the reciprocal of the Rossby number $\mathrm{Ro}$ (defined by equation \ref{eq:def_rossby}) and $\Delta \Omega_\mathrm{core} / \Omega_\mathrm{eq}$, respectively. 
    The circles with error-bars represent the observational values obtained from \citet{Hall2021NatAs}, \citet{Benomar2018Sci} and \citet{Bazot2019AandA}. The blue filled circles with texts are the solar-analogue stars, and the blue mark $\odot$ is the solar value. The triangles represent the numerical result of \citet{Brun2022ApJ}. 
    Note that we limit the mass range to $0.85-1.25 \, \mathrm{M}_\odot$ for both observational values and numerical results. The colours of the solid lines correspond to $\alpha_\mathrm{ef}$ shown in the colour bar. The black dashed line shows the result of the rigid-body rotation with $\Delta \Omega_\mathrm{core}=0$.
    }
    \label{fig:DRmodel}
\end{figure}

Fig.~\ref{fig:DRmodel} compares the normalised LDR, $\Delta \Omega_\mathrm{core}/ \Omega_\mathrm{eq}$, derived from equation (\ref{eq:Omega_Rossby_formula}) for different values of $0.05\le\alpha_\mathrm{ef}\le 0.60$ (solid lines) to the numerical simulation
    \footnote{$\Delta \Omega_\mathrm{core}$ of the numerical simulation is directly read from the data of \citet{Brun2022ApJ} supplied by Alan Sacha Brun and Antoine Strugarek. We use the data of $M = 0.9,1.1 \, \mathrm{M}_\odot$ for consistency to the observed mass range.}
\citep[filled triangle; ][]{Brun2022ApJ} and observations \citep[circles with error bars; ][]{Hall2021NatAs, Benomar2018Sci,Bazot2019AandA}. The observed data are listed in Table~\ref{tab:landscape} in Appendix \ref{app:SampleSelection}. It should be noted that the only 13 samples out of 40 observed stars by \citet{Benomar2018Sci} are plotted in Fig.~\ref{fig:DRmodel}; in reality, the remaining 27 samples that show no clear evidence of LDR are possibly distributed near $\Delta \Omega_\mathrm{core} /  \Omega_\mathrm{eq} \sim 0$ 
\footnote{It is possible that they have substantial LDR, although the probability is expected to be lower than the samples whose LDR are detected.}
. The presented observational data exhibit a large scatter in LDR; even `solar-analogue' stars (blue circles) with Ro $\approx$ Ro$_\mathrm{\odot}$ show different levels of LDR. In particular, KIC9025370 shows a quite strong LDR, $\Delta \Omega_\mathrm{core} /  \Omega_\mathrm{eq} > 0.5  $, which can be explained by a large value of $\alpha_\mathrm{ef} >0.5$ in our model. These solar analogue stars suggest the scenario that the Sun had strong equator-fast DR in the past and is now experiencing a rapid suppression of the LDR. We discuss this scenario in Section. \ref{sec:Discussion}.

\subsection{Wind Torque Model} \label{subsec:Wind_model}

Wind torque is directly linked to the mass-loss rate by magnetised stellar winds, $\dot{M}$, and the Alfv\'{e}n radius, $r_{\rm A}$, via 
\begin{equation}
    \tau_\mathrm{w} \approx \dot{M} r_\mathrm{A}^2 \Omega_\mathrm{ave} ,
\label{eq:torque_approx}
\end{equation} 
where $r_\mathrm{A}$ is defined as the location of the cylindrical surface where the radial wind velocity equals the radial component of the local Alfv\'{e}n speed \citep[e.g.][]{Weber1967ApJ}. \citet{Matt2012ApJ} conducted a large number of MHD simulations to derive a scaling relation of the wind torque:
\begin{equation}
    \tau_\mathrm{w} \propto M^{-\ell} R^{5\ell+2} B^{4\ell} \dot{M}^{1-2\ell} \Omega_\mathrm{ave}, \label{eq:scaling_tau_w}
\end{equation}
where $M$, $R$ and $B$ denote the stellar mass, the stellar radius, and the magnetic field strength of the dipole component \citep[][]{Finley2018ApJ_Geo2} strength on the surface, respectively. We note that previous numerical simulations give a small value of the exponent factor $\ell = 0.20$-$0.25$ \citep[see][]{Matt2008ApJ,Matt2012ApJ,Finley2018ApJ_Geo2}. The relatively strong dependencies on $R$ and $B$ in equation (\ref{eq:scaling_tau_w}) mainly arise from those of $r_{\rm A}$. In contrast, $\tau_{\rm w}$ depends only weakly on $\dot{M}$ because the negative correlation of $r_{\rm A}$ with $\dot{M}$ partially cancels the linear dependence on $\dot{M}$ in equation (\ref{eq:torque_approx}) \citep{Washimi1993MNRAS,Shoda2020ApJ}.

Observations using Zeeman-Doppler imaging confirmed a positive dependence of $B$ on stellar rotation rate \citep{Vidotto2014MNRAS,See2019ApJ}. Numerical simulations for Alfv\'{e}n wave-driven magneto-rotating winds also gave a moderately positive dependence of mass-loss rate on rotation rate \citep{Shoda2020ApJ}. In addition to these tendencies, $B$ and $\dot{M}$ are expected to be dependent on stellar mass and radius. Incorporating all these dependencies, one can rewrite $\tau_{\rm w}$ in equation (\ref{eq:scaling_tau_w}) as a function of the only three basic parameters, $M$, $R$, and $\Omega_{\rm ave}$.

\citet{Matt2015ApJ} introduced such a simple formulation, which we follow in this paper, namely, we 
parameterise the wind torque as 
\begin{equation}
    \tau_\mathrm{w} \equiv \tau_\mathrm{w,\odot} X_1 (R,M) X_2 (\Omega_{\rm core},\mathrm{Ro}, \Omega_{\rm eq}). \label{eq:def_tau_w}
\end{equation}
We note that the dependencies are separated into those on the mass and radius, $X_1$, and those on the rotational properties, $X_2$.

We adopt the same function as in \citet{Matt2015ApJ} for $X_1 (R,M)$ :
\begin{equation}
    X_1 (R,M) \equiv \left( \dfrac{R }{\mathrm{R}_\odot} \right)^{u} \left( \dfrac{M }{\mathrm{M}_\odot} \right)^{v},
\label{eq:def_X1}
\end{equation}
where $u=3.1$ and $v=0.5$, which explain well the observed rotation rates of different-mass stars \citep{Matt2015ApJ}. Comparison between equations (\ref{eq:scaling_tau_w}) and (\ref{eq:def_X1}) indicates that the relation of $B^{4\ell} \dot{M}^{1-2\ell} \propto R^{u-(5\ell+2)} M^{v+\ell} X_2(\Omega_{\rm core},\mathrm{Ro}, \Omega_{\rm eq})$ is assumed. 
In addition to $M$ and $R$, metallicity can also be an additional factor that affects the mass-loss rate \citep{Suzuki2018PASJ}, and accordingly, the wind torque (Section \ref{subsec:D_metal}). However, we do not consider this effect because it is regarded to be subdominant, compared to the effect of the modification in the stellar structure due to a difference in metallicity \citep[][see also the discussion in Section \ref{subsec:SDE_metaldepend} and Appendix \ref{app:MESA}]{Amard2020ApJ}.

$X_2(\Omega_{\rm core},\mathrm{Ro}, \Omega_{\rm eq})$ denotes the torque regulated by the stellar rotation, including LDR. An important update from \citet{Matt2015ApJ} is that we explicitly include the dependence of $X_2$ on $\Omega_{\rm eq}$ in addition to $\Omega_\mathrm{core}$ and $\mathrm{Ro}$ in order to consider the LDR effect. We note that $\Omega_\mathrm{eq}=\Omega_\mathrm{core}+\Delta \Omega_\mathrm{core}$ (equation \ref{eq:def_Delta_Omega}) is analytically calculated from $\Omega_\mathrm{core}$ and $\mathrm{Ro}$ via equation (\ref{eq:Omega_Rossby_formula}). (see Section \ref{subsec:DR_model}). The specific form of $X_2(\Omega_{\rm core},\mathrm{Ro},\Omega_{\rm eq}(\Omega_{\rm core},\mathrm{Ro}))$ is given by
\begin{equation}
    X_2(\Omega_{\rm core},\mathrm{Ro}, \Omega_{\rm eq}) \equiv \begin{cases}
    & \left( \dfrac{\mathrm{Ro} }{\mathrm{Ro}_\odot} \right)^{-p} \left( \dfrac{\Omega_{\rm core }}{\Omega_{\rm core,\odot}} \right)^{1-q} \left( \dfrac{\Omega_{\rm eq }}{\Omega_{\rm eq,\odot}} \right)^{q} \\[3mm] 
    & \hspace{10pt} \text{for $\mathrm{Ro}  > \mathrm{Ro}_{\rm crit}$ (Slow \& Moderate)}, \\[3mm]
    &\chi^{p} \left( \dfrac{\Omega_{\rm core }}{\Omega_{\rm core,\odot}} \right)^{1-q} \left( \dfrac{\Omega_{\rm eq }}{\Omega_{\rm eq,\odot}} \right)^{q} \\[3mm]
    & \hspace{10pt} \text{for $\mathrm{Ro} \leq \mathrm{Ro}_{\rm crit}$ (Fast)}, 
\end{cases}
\label{eq:def_X2}
\end{equation}
where $\chi \equiv (\mathrm{Ro}_{\rm crit} / \mathrm{Ro}_\mathrm{\odot})^{-1}$ and $p$ and $q$ are the scaling indices about $\mathrm{Ro}$ and $\Omega_\mathrm{eq}$. These two branches correspond to unsaturated and saturated stars with respect to magnetic activity \citep[][; see also Section \ref{subsec:DR_model}]{Wright2011ApJ,See2019ApJ}. Note that the original formulation of \citet{Matt2015ApJ} without LDR is recovered when $q=0$.

Using equation (\ref{eq:def_rossby}), we rewrite the unsaturated branch (slow \& moderate rotators) of equation (\ref{eq:def_X2}) into 
\begin{equation}
    X_2 =
    \left( \dfrac{\tau_\mathrm{cz}^\mathrm{CS} }{\tau_\mathrm{cz,\odot}^\mathrm{CS}} \right)^{p} \left( \dfrac{\Omega_{\rm core }}{\Omega_{\rm core,\odot}} \right)^{p+1-q} \left( \dfrac{\Omega_{\rm eq }}{\Omega_{\rm eq,\odot}} \right)^{q} \text{for $\mathrm{Ro}  > \mathrm{Ro}_{\rm crit}$}.
\label{eq:re_X2}
\end{equation}
Substituting equation (\ref{eq:re_X2}) into equation (\ref{eq:def_tau_w}) and assuming no LDR ($\Omega_{\rm eq} = \Omega_{\rm core} = \Omega$), one may find the scaling relation of $\tau_w \propto \Omega^{1+p}$ introduced in Section \ref{sec:intro}. Throughout this paper, we fix $p=2$, which corresponds to the Skumanich law. The index $q$ governs the LDR effect in determining the wind torque. We treat $q$ in a range between $0$ and $p+1(=3)$. Note that $q=p+1(=3)$ is the extreme case in which the magnetic braking is completely controlled by the angular velocity at the equator, while in the case of $q=0$ it is determined by the angular velocity of the core.

\begin{figure}
    \centering
	\includegraphics[width=\columnwidth]{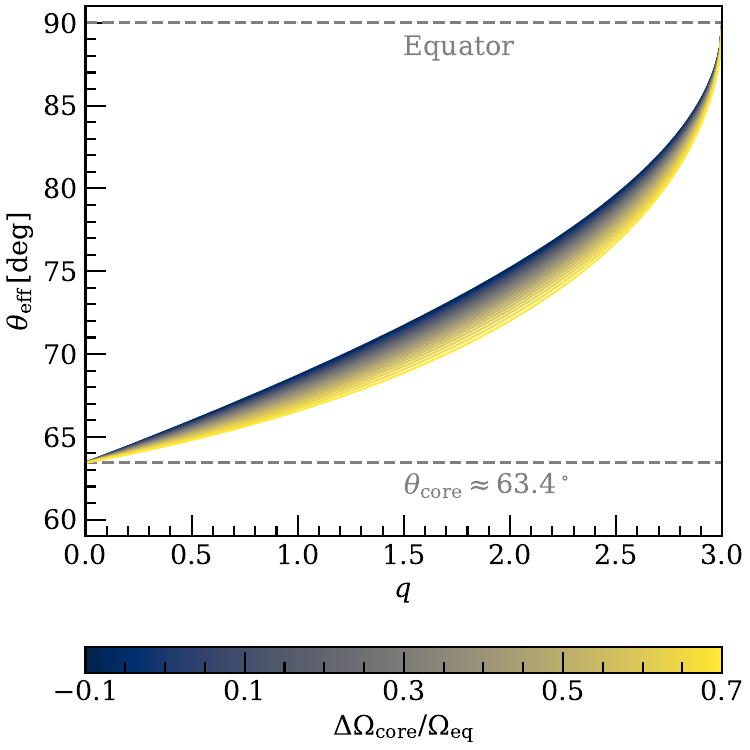}
    \caption{The relation between $q$ and $\theta_\mathrm{eff}$ for different values of $\Delta \Omega_\mathrm{core}$ (the lower colour bar). The abscissa and ordinate represent the strength of the LDR effect $q$ and the effective co-latitude  $\theta_\mathrm{eff}$, respectively. The upper horizontal dashed line corresponds to $\theta_\mathrm{eff}=90^{\circ}$, which is the equator of the star. The lower horizontal dashed line corresponds to $\theta_\mathrm{eff}=\theta_\mathrm{core}\approx 63.4^{\circ}$ (the definition is given in the texts after equation \ref{eq:def_Delta_Omega}), which satisfies $\Omega_\mathrm{env}(\theta_\mathrm{core})=\Omega_\mathrm{core}$. }
    \label{fig:q_thetaeff}
\end{figure}

The physical meaning of $q$ is quantitatively clearer if we introduce the effective co-latitude, $\theta_\mathrm{eff}$  \citep[cf.][]{Ireland2022ApJ}, that satisfies
\begin{gather}
    \left( \dfrac{ \Omega_\mathrm{env}(\theta_\mathrm{eff})}{\Omega_\mathrm{env,\odot}(\theta_\mathrm{eff})} \right)^{p+1} = \left( \dfrac{ \Omega_\mathrm{core}}{\Omega_\mathrm{core,\odot}} \right)^{p+1-q} \left( \dfrac{ \Omega_\mathrm{eq}}{\Omega_\mathrm{eq,\odot}} \right)^{q}.
    \label{eq:def_theta_eff}
\end{gather}
Comparing equations (\ref{eq:re_X2}) and (\ref{eq:def_theta_eff}), one may understand that $\theta_{\rm eff}$ is the characteristic co-latitude that determines the wind torque via equation (\ref{eq:def_tau_w}). Fig.~\ref{fig:q_thetaeff} clearly shows that $\theta_{\rm eff}$ monotonically increases with $q$ from $\theta_{\rm eff}=\theta_{\rm core}$ for $q=0$ to $\theta_{\rm eff}=90^{\circ}$ for $q=3$. In particular, $67^{\circ} \lesssim \theta_\mathrm{eff} \lesssim 75^{\circ}$ when $1 \leq q \leq 2$, regardless of the value of $\Delta \Omega_\mathrm{core}$. Interestingly, this range is consistent with the opening co-latitude $\sim 70^{\circ}$ of the solar wind during solar maximum \citep{Ireland2022ApJ,FinleyBrun2023arXiv}. Bearing in mind that active solar-type stars are in some sense an extension of the Sun during the high-activity phase \citep[e.g.,][]{Telleschi2005ApJ,Gudel2007LRSP}, we suppose $1 \leq q \leq 2$ is a reasonable range for our study. On the flip side, while $q$ may vary with time in realistic situations, we assume a constant $q$ in this paper for simplicity.

\subsection{The AM evolution model}
\label{subsec:AM_evo}
The long-term evolution of stellar AM can be calculated by equation (\ref{eq:AMev_intro}). In our setup, the magnetic braking primarily works on the low-latitude region in the envelope (Sections \ref{subsec:picture} and \ref{subsec:Wind_model}). Therefore, in principle,  the stellar rotation is decelerated most severely there. Here, we assume that the redistribution of the AM inside the star takes place with a sufficiently short timescale in comparison to the spin-down, so that the internal profile of $\Omega(r,\theta)$ (equation \ref{eq:DRprofile_model} with equation \ref{eq:Omega_Rossby_trend}) is preserved with decreasing $\Omega_{\rm core}$. We note that this assumption corresponds to the case of zero coupling time in the two-layer model of \citet{MacGregor1991ApJ}.

Under this assumption and equation (\ref{eq:def_AM}), we rewrite equation (\ref{eq:AMev_intro}) into 
\begin{equation}
    \dfrac{\mathrm{d} \Omega_\mathrm{core}}{\mathrm{d}t}=- \dfrac{\tau_\mathrm{w}}{I} - \dfrac{\Omega_\mathrm{core}}{I} \frac{\mathrm{d} I}{\mathrm{d}t}. \label{eq:AM_eq}
\end{equation}
Here, $I(t)$ as well as $M(t)$, $R(t)$ and $T_\mathrm{eff}(t)$ is determined by the \textsc{mesa} stellar evolution code \citep[r12778,][]{paxton2011modules, paxton2013modules, paxton2015modules, paxton2018modules, paxton2019modules}.  The details of parameter settings of \textsc{mesa} and the obtained profiles of $R(t),I(t)$, and $\tau_\mathrm{cz}^\mathrm{CS}(T_\mathrm{eff}(t))$ are shown in Appendix \ref{app:MESA}. We set the stellar mass to $M=0.9, 1.0, 1.1$ and $1.2 \, \mathrm{M}_{\odot}$. 

We consider three different initial metallicities, $Z=0.5, 1.0, 2.0 \, \mathrm{Z_\odot}$, where $Z$ is the mass fraction of the heavy elements. The solar abundances are adopted from \citet[][]{GS1998SSRv}, who give the mass fraction of helium, $\mathrm{Y_\odot}=0.270$, and the heavy elements, $\mathrm{Z_\odot}=0.018$. The initial helium abundance is given by the enrichment law $(Y-1.0 \, \mathrm{Y_\odot}) = 2.0 \times (Z-1.0 \, \mathrm{Z_\odot})$ \citep[e.g.][]{Chiosi1982A&A}. We set the initial rotation frequency ($\Omega_\mathrm{core,init}$) to the range of  $2.7 \, \Omega_\mathrm{core,\odot} \leq \Omega_\mathrm{core,init} \leq 27 \, \Omega_\mathrm{core,\odot}$ at the zero-age main sequence (ZAMS). We note that, in terms of rotation period $P_\mathrm{rot}$, this range of angular frequency corresponds to $1 \, \mathrm{d} \lesssim P_\mathrm{rot} \lesssim 10 \, \mathrm{d}$, which is the same condition as that  of \citet{Matt2015ApJ}. We set the default initial rotation frequency to $\Omega_\mathrm{core,init} = 10 \, \Omega_\mathrm{core,\odot}$ (see Section \ref{sec:Result} for the dependence on $\Omega_\mathrm{core,init}$). By solving equation (\ref{eq:AM_eq}) with the wind torque (equation \ref{eq:def_tau_w}) that explicitly considers the effect of LDR (equation \ref{eq:def_X2}), we calculate the time evolution of $\Omega_{\rm core}$ from the ZAMS.

\begin{figure}
    \centering
	\includegraphics[width=1.0\columnwidth]{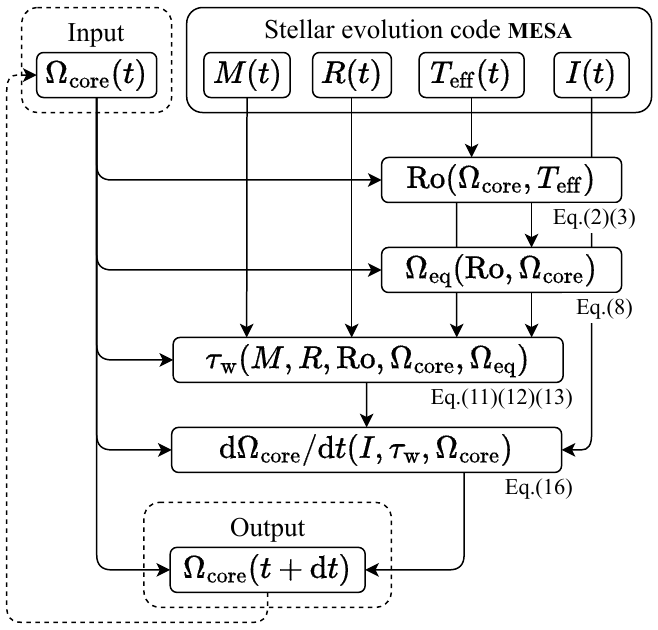}
    \caption{The flowchart of our calculation,  which is constructed from equations  (\ref{eq:def_rossby}), (\ref{eq:timescale_overturn}), (\ref{eq:Omega_Rossby_formula}),
    (\ref{eq:def_tau_w}), 
    (\ref{eq:def_X1}),
    (\ref{eq:def_X2})
    and
    (\ref{eq:AM_eq}) and the stellar evolution code \textsc{mesa}.}
    \label{fig:flowchart}
\end{figure}

Fig.~\ref{fig:flowchart} shows the flowchart of our calculation. Table \ref{tab:parameter} lists the adopted parameters.

\subsection{Expected effects of LDR on stellar spin-down} 
\label{subsec:SDE_trend}

\begin{figure}
    \centering 
    \includegraphics[width=0.95\columnwidth]{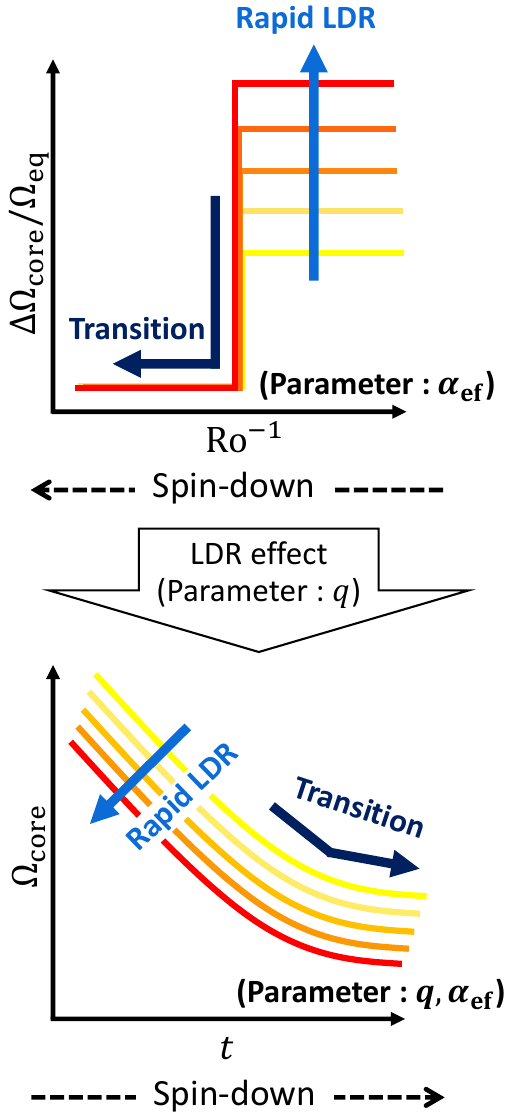}
    \caption{The schematic picture of how the evolution of LDR (upper panel) affects the spin-down (lower panel). The top diagram is a schematic representation of Fig.~\ref{fig:DRmodel}; the bottom diagram illustrates the resultant evolution of $\Omega_\mathrm{core}$; here the colour of the solid lines corresponds to each other in both diagrams. Note that the direction of the evolution is leftward (rightward) in the top (bottom) diagram (dashed arrows).
    When stronger LDR is considered (light blue arrow in top diagram), stellar rotation slows down more effectively from earlier times (light blue arrow in bottom diagram). The transition from equator-fast DR to pole-fast DR (dark blue arrow in top diagram) yields the concave up profile of $\Omega_\mathrm{core}(t)$ (dark blue arrow in bottom diagram). The variables with parentheses denote the control parameters in each diagram. 
    }
    \label{fig:schematic_trend}
\end{figure}

Before presenting solutions of $\Omega_\mathrm{core}(t)$ for various sets of $(\alpha_\mathrm{ef}, q, M, \Omega_\mathrm{core,init})$, let us illustrate how the time evolution of stellar rotation is modified by the effect of LDR with schematic diagrams (Fig.~\ref{fig:schematic_trend}). The characteristic modifications  are 1. the magnetic braking is enhanced in the early phase when a star is rotating with strong equator-fast DR and 2. the transition from equator-fast DR to pole-fast DR weakens the magnetic braking in the late phase. As a result, the evolutionary track of $\Omega_\mathrm{core} (t)$ is concave up in a $t - \Omega_\mathrm{core}$ plane (bottom diagram of Fig.~\ref{fig:schematic_trend}). A byproduct of modification 1. is that the observed trend of solar-type stars can be explained by a smaller solar wind torque than the value in the case without LDR effect \citep[e.g.][]{Matt2015ApJ} A direct outcome of modification 2. is that we can reproduce the observed trend of weakened magnetic braking  \citep[e.g.][]{vanSaders2016Nature} (see Section \ref{sec:intro}).

\begin{table}
	\centering
	\caption{Adopted parameter values.}
        \setlength\tabcolsep{0.05cm}
	\label{tab:parameter}
	\begin{tabular}{cll} 
		\hline
		Symbol & Adopted Value & Description  \\
		\hline
  
        $\mathrm{M}_\odot$ & $1.99 \times 10^{33} \, \mathrm{g}$ & Solar mass  \\
        
        $\mathrm{R}_\odot$ &  $6.96 \times 10^{10} \, \mathrm{cm}$ & Solar radius  \\
        
        $\mathrm{I}_{\odot}$ &  $7.13 \times 10^{53} \, \mathrm{g \cdot cm^2 }$ & Solar moment of inertia  \\
        $\mathrm{T}_{\mathrm{eff},\odot}$ &  $5.77 \times 10^{3} \, \mathrm{K}$ & Solar effective temperature  \\
        
        $t_\odot$ & $4.57 \times 10^{10} \, \mathrm{Gyr}$ & Solar age  \\
        
        $\mathrm{Z}_\odot$ & 0.018 & Solar initial mass fraction of heavy-element \\
        
        $\mathrm{Y}_\odot$ & 0.270 & Solar initial mass fraction of helium \\
        
        $\Omega_\mathrm{core,\odot}$ & $ 436.5 \, \mathrm{nHz / 2\pi}$ & Solar core rotation rate  \\
        
        $\Delta \Omega_\mathrm{core,\odot}$ & $ 15.0 \, \mathrm{nHz / 2\pi}$ & Solar LDR extent (see equation \ref{eq:def_Delta_Omega}).  \\
        
        $\tau^{\rm CS}_{\rm cz,\odot}$ & $1.29 \times 10 \, \mathrm{d}$ & Solar convection turnover time obtained \\
        & & by Equation  (\ref{eq:timescale_overturn}). \\
        
        $\chi$ & 10 & The Reciprocal of critical Rossby number \\
        & & $\mathrm{Ro_{crit}}^{-1}$ normalised by the solar value $\mathrm{Ro}_\odot^{-1}$  \\
        
        $n,m$ & 20 & The smoothness indices introduced in \\
        & & equation (\ref{eq:Omega_Rossby_formula}). \\
        
        $p$ & 2 & The scaling index of $\tau_\mathrm{w} \propto \Omega^{1+p}$ which \\
        & & conforms to Skumanich's law  \\
		\hline
	\end{tabular}
\end{table}

\section{Observational data} 
\label{sec:observation}

In this section, we introduce properties of observational data briefly. We adopt the stellar age achieved from the asteroseismic modelling \citep{Hall2021NatAs}. Observed $\Omega_\mathrm{core}$ is obtained from two different methods: rotational modulation of radiation flux by star spots \citep{Garcia2014AandA, Lu2022ApJ} and rotational splitting of eigen-frequencies of stellar oscillations \citep{Benomar2018Sci,Bazot2019AandA}. While the latter method can capture the rotational property of the core with high accuracy thanks to the asteroseismic technique, the former method may suffer systematic errors because what is measured is the rotation of star spots appearing on the surface (see discussion in Section \ref{subsec:D_errbiasobs}). We adopt the asteroseismic values (spot-modulation-based values) for the stars whose LDR are detected (not-detected) distinctly. Our sample selection is provided in detail in Appendix \ref{app:SampleSelection}.

\section{Results} \label{sec:Result}

\subsection{Evolution of 1 $\mathrm{M}_\odot$ stars} \label{subsec:SDE_1Msun}

\begin{figure*}
    \centering
	\includegraphics[width=2.0\columnwidth]{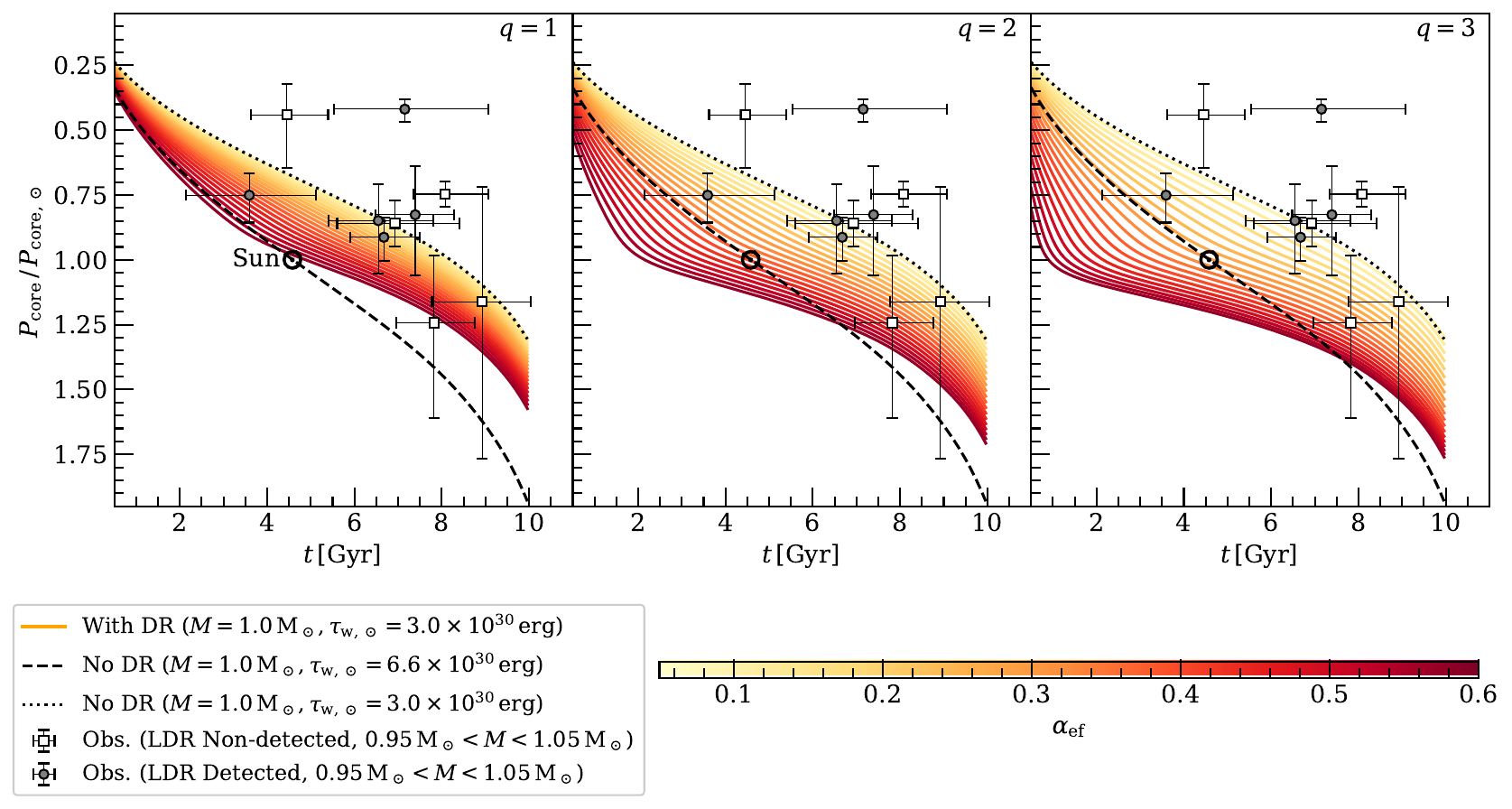}
    \caption{The spin-down of the stars with $M= 1.0 \, \mathrm{M}_\odot$ and $Z=1.0 \, \mathrm{Z}_\odot$. We set the initial core angular velocity to $\Omega_\mathrm{core} = 10 \, \Omega_\mathrm{core,\odot}$. The abscissa and ordinate represent the time and the rotation period of the core normalised by the solar value, respectively. The left, middle, and right panels correspond to the case of $q=1,2,3$, respectively. The coloured solid line shows the result of our model when $\tau_\mathrm{w,\odot}=3.0 \times 10^{30} \, \mathrm{erg} \approx \tau_\mathrm{w,\odot,F19}$ \citep[][]{Finley2019ApJ_Direct}, where the colours correspond to $\alpha_\mathrm{ef}$ shown in the colour bar. The black dashed (dotted) line shows the result under the rigid-body rotation when   $\tau_\mathrm{w,\odot}=6.6 \times 10^{30} \, \mathrm{erg} \approx \tau_\mathrm{w,\odot,M15}$  ($\tau_\mathrm{w,\odot}=3.0 \times 10^{30} \, \mathrm{erg} \sim \tau_\mathrm{w,\odot,F19}$). The filled circles (open squares) with error-bars represent observational values of stars with $0.95\, \mathrm{M}_\odot < M < 1.05 \, \mathrm{M}_\odot$ whose LDR are detected (non-detected) distinctly (see Section \ref{sec:observation} and Appendix \ref{app:SampleSelection}). The mark `$\odot$' represents the solar value.
    }
    \label{fig:SDE_1p0M}
\end{figure*}

\begin{figure}
    \centering
	\includegraphics[width=0.8\columnwidth]{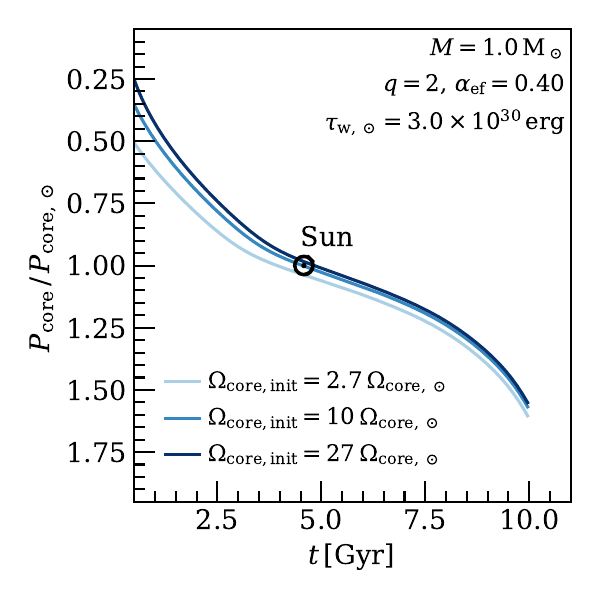}
    \caption{The spin-down of the stars with $M= 1.0 \, \mathrm{M}_\odot$ and $Z=1.0 \, \mathrm{Z}_\odot$ for three different cases of $\Omega_\mathrm{core,init}$ $= 2.7\, \Omega_\mathrm{core,\odot}$ (light blue), $10\, \Omega_\mathrm{core,\odot}$ (middle blue), and $27\, \Omega_\mathrm{core,\odot}$ (dark blue).
    We set $q=2$, $\alpha_\mathrm{ef}=0.40$, and $\tau_\mathrm{w,\odot}=3.0 \times 10^{30} \, \mathrm{erg}$. The abscissa and ordinate are the same as those in Fig.~\ref{fig:SDE_1p0M}. 
    }
    \label{fig:SDE_init}
\end{figure}

\begin{figure}
    \centering
	\includegraphics[width=\columnwidth]{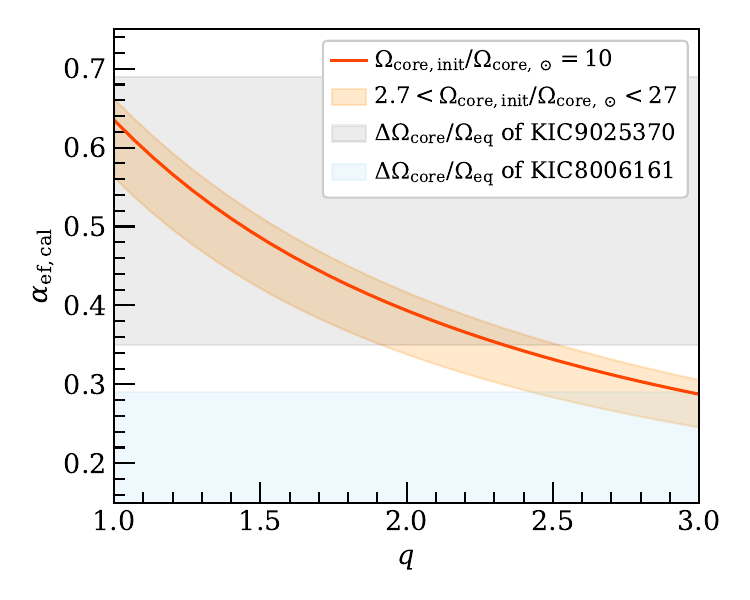}
    \caption{ The value of $\alpha_\mathrm{ef,cal}(q)$ that gives the current solar rotation, $\Omega_\mathrm{core}(t_\odot)=\Omega_\mathrm{core,\odot}$. The orange line is for the initial angular velocity, $\Omega_\mathrm{core,init} = 10 \, \Omega_\mathrm{core,\odot}$. The orange shaded region is the range for $2.7 \, \Omega_\mathrm{core,\odot} < \Omega_\mathrm{core,init} < 27 \, \Omega_\mathrm{core,\odot}$ \citep{Matt2015ApJ} , whereas the upper and lower boundaries correspond to $27 \, \Omega_\mathrm{core,\odot}$ and $2.7 \, \Omega_\mathrm{core,\odot}$, respectively. The grey and blue shaded region represents the observed constraint derived from KIC9025370 and KIC8006161, respectively (see text and Fig.~\ref{fig:DRmodel}). 
    }
    \label{fig:q_alphaefcal}
\end{figure}

\begin{figure*}
    \centering
	\includegraphics[width=2.0\columnwidth]{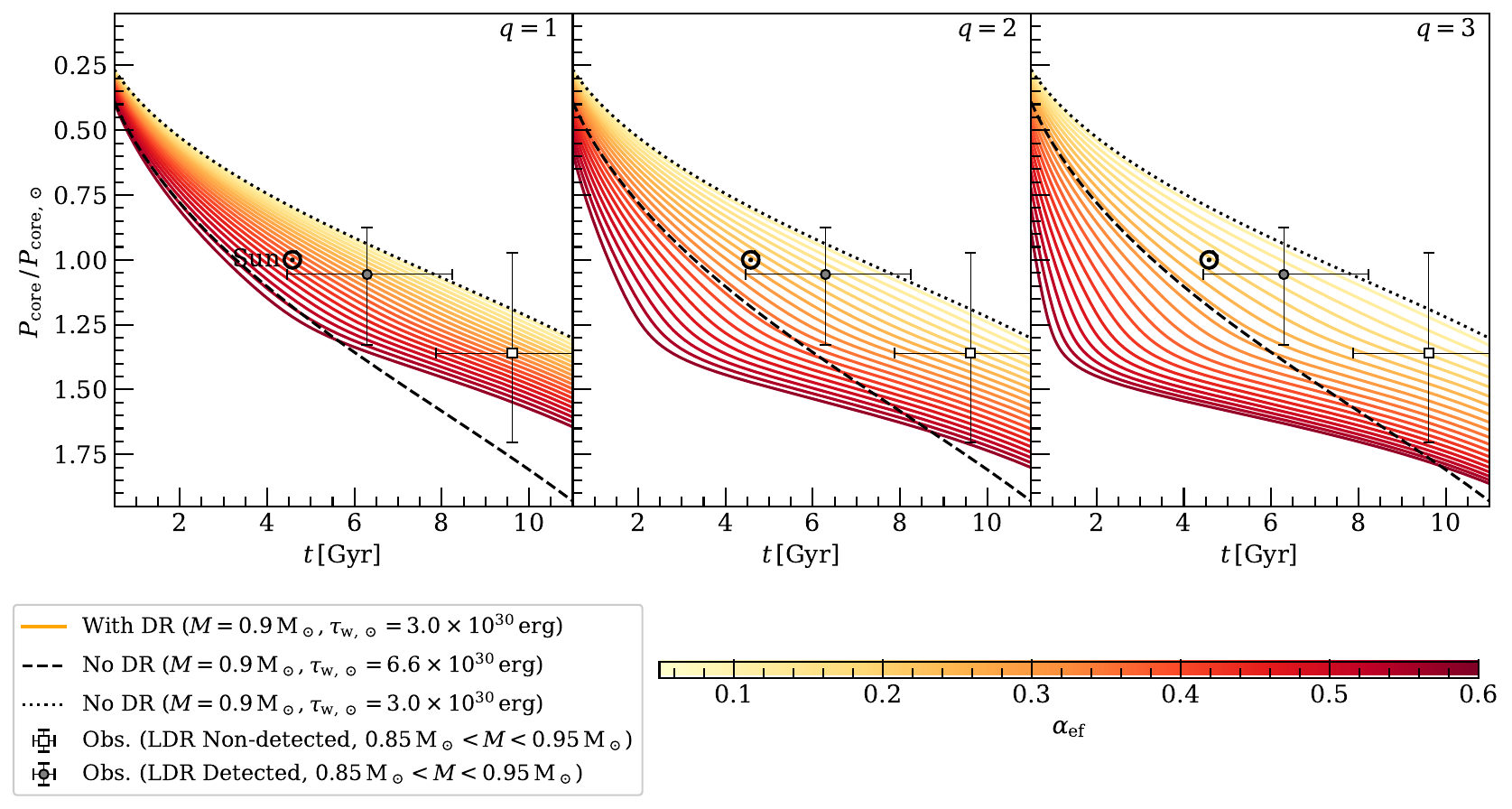}
    \caption{Same as Fig.~\ref{fig:SDE_1p0M} but for models at $M= 0.9 \, \mathrm{M_\odot}$. The observational data are taken from the stars with $0.85\, \mathrm{M}_\odot < M < 0.95 \, \mathrm{M}_\odot$. }
    \label{fig:SDE_0p9M}
\end{figure*}

\begin{figure*}
    \centering
	\includegraphics[width=2.0\columnwidth]{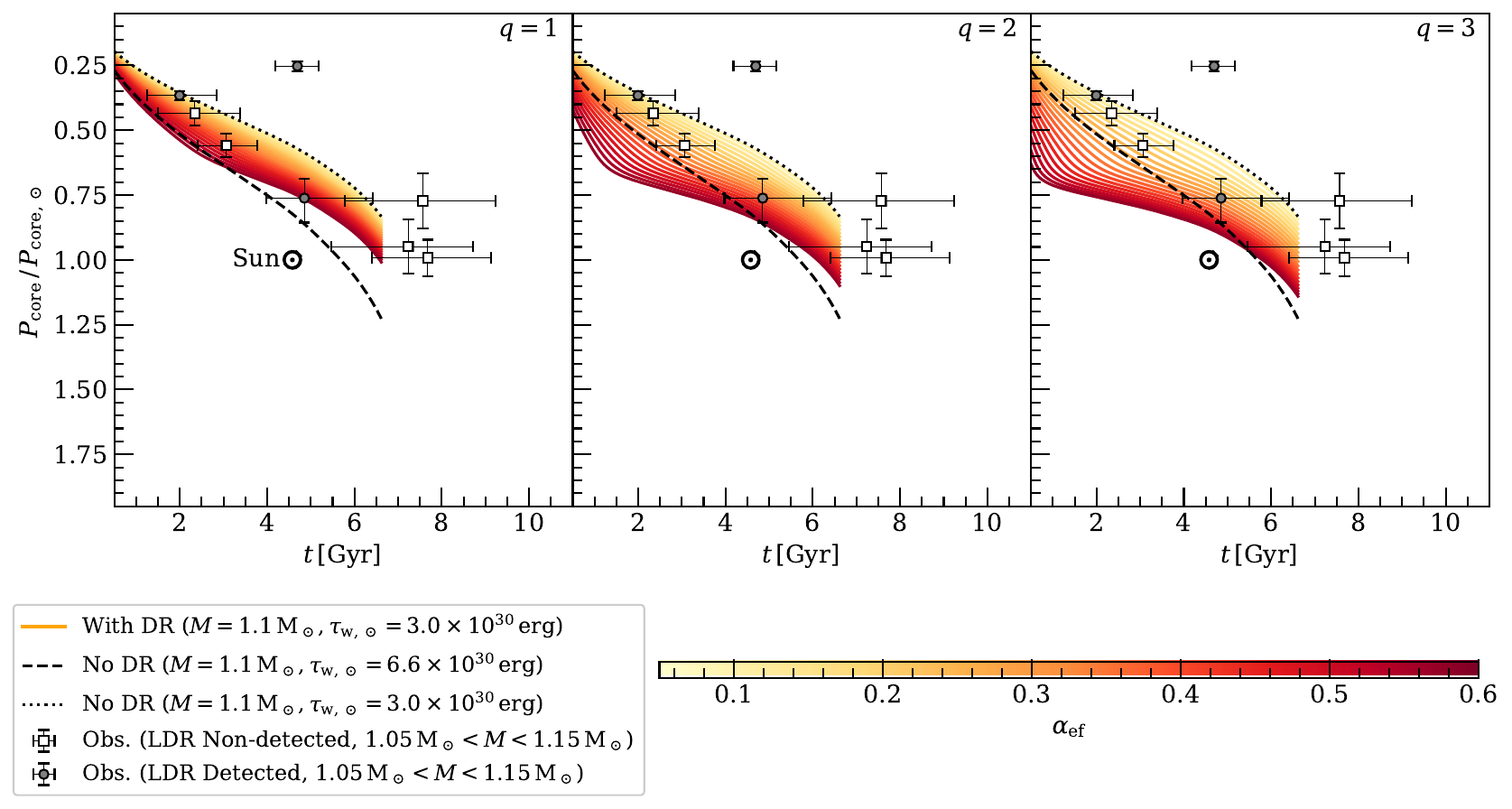}
    \caption{Same as Fig.~\ref{fig:SDE_1p0M} but for models at $M= 1.1 \, \mathrm{M_\odot}$.The observational are from the stars with $1.05\, \mathrm{M}_\odot < M < 1.15 \, \mathrm{M}_\odot$. }
    \label{fig:SDE_1p1M}
\end{figure*}

\begin{figure*}
    \centering
	\includegraphics[width=2.0\columnwidth]{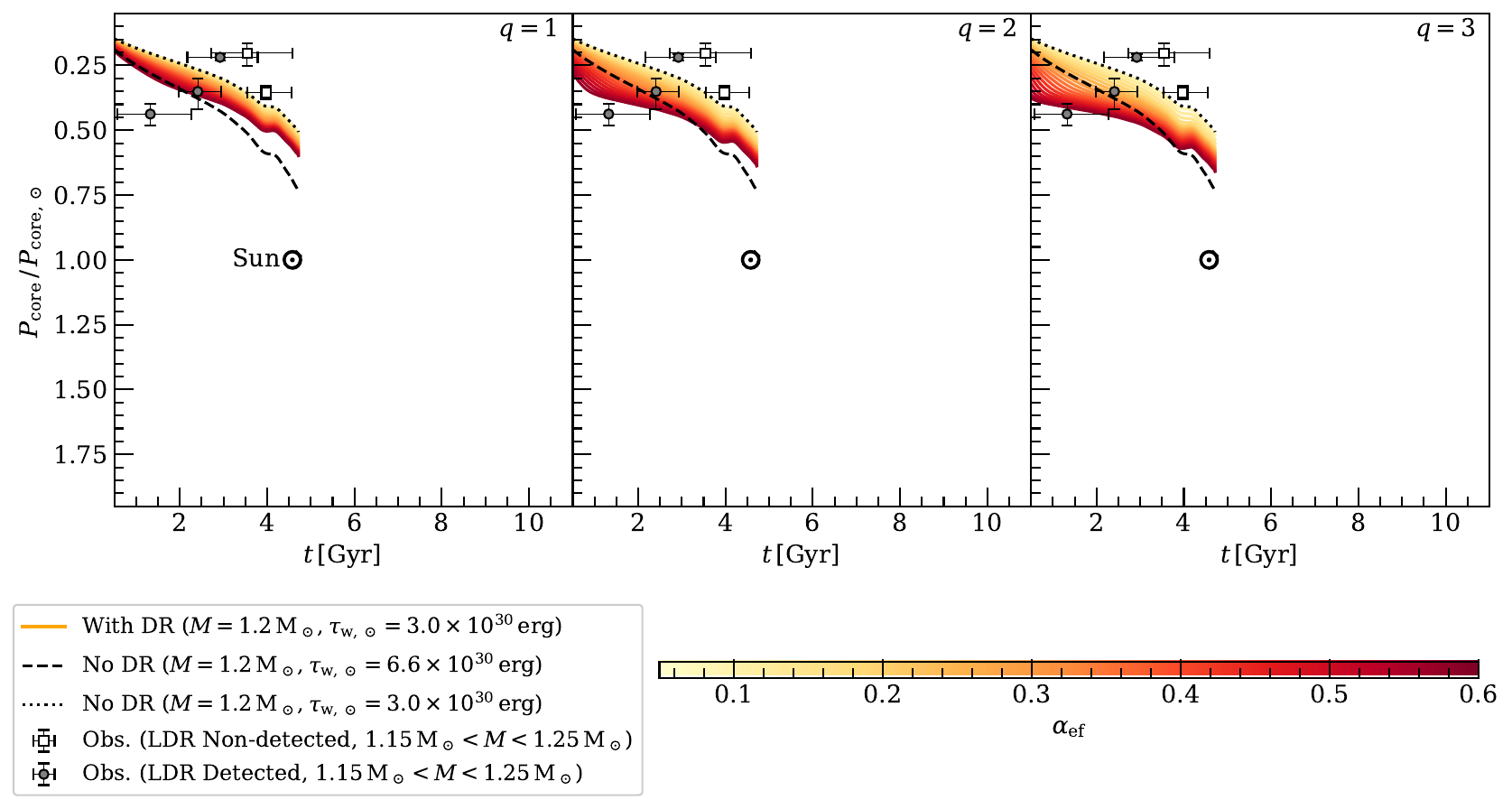}
    \caption{Same as Fig.~\ref{fig:SDE_1p0M} but for models at $M= 1.2 \, \mathrm{M_\odot}$.The observational are from the stars with $1.15\, \mathrm{M}_\odot < M < 1.25 \, \mathrm{M}_\odot$. }
    \label{fig:SDE_1p2M}
\end{figure*} 

\begin{figure}
    \centering
	\includegraphics[width=0.8\columnwidth]{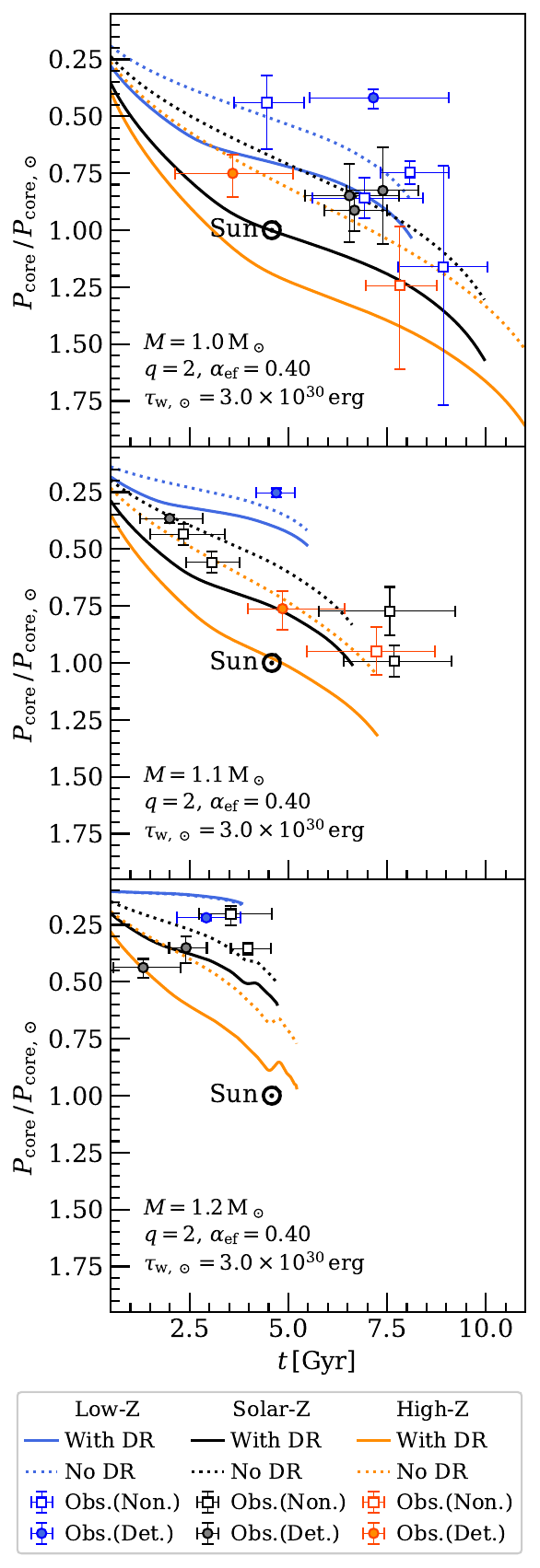}
    \caption{
    The spin evolution of the stars with $M= 1.0 \, \mathrm{M}_\odot$ (top panel), $M= 1.1 \, \mathrm{M}_\odot$ (middle panel) and $1.2 \, \mathrm{M}_\odot$ (bottom panel) for three cases of the stellar abundance: $Z=0.5 \, \mathrm{Z}_\odot$ (blue lines), $1.0 \, \mathrm{Z}_\odot$ (black lines), and $2.0 \, \mathrm{Z}_\odot$ (orange lines).  The solid (dotted) lines represent the model with (without) the LDR effect when $\tau_\mathrm{w,\odot}=3.0 \times 10^{30} \, \mathrm{erg}$. For the parameters on the LDR effect, we adopt $q=2$, $\alpha_\mathrm{ef}=0.40$. 
    The abscissa, ordinate and observations are the same as those in Figs.~\ref{fig:SDE_1p0M},\ref{fig:SDE_1p1M}, and \ref{fig:SDE_1p2M} but the colour of the observational data denotes the stellar metallicity: blue, black, and orange correspond to metallicity ranges of $\mathrm{[Fe/H]} < -0.15$, $-0.15 \leq \mathrm{[Fe/H]} \leq 0.15$, $0.15 < \mathrm{[Fe/H]}$, respectively.%
    }
    \label{fig:SDE_metal}
\end{figure}

Fig.~\ref{fig:SDE_1p0M} presents the spin  evolution of the stars with $M= 1.0 \, \mathrm{M}_\odot$ with (coloured lines) and without (black lines) the LDR effect for various $q$ and $\alpha_\mathrm{ef}$ values, in comparison to the observed solar-like stars with $0.95 \, \mathrm{M}_\odot < M < 1.05 \, \mathrm{M}_\odot$ \citep[circles and squares with error bars;][see Table \ref{app:SampleSelection} in Appendix  \ref{app:SampleSelection} for the detailed properties of the observed data]{Garcia2014AandA, Benomar2018Sci,Bazot2019AandA,Hall2021NatAs,Lu2022ApJ}. The cases with the LDR effect adopt $\tau_{\rm w,\odot}=3.0\times 10^{30}\, \mathrm{erg} = \tau_\mathrm{w,\odot,F19}$ obtained from the direct measurement in the solar wind by WIND and PSP \citep[][]{Finley2019ApJ_Direct, Finley2020ApJ} for the normalisation of $\tau_\mathrm{w}$. Comparing the coloured lines with the black dotted lines, we find that the stellar rotation is significantly decelerated from the early phase by including the LDR effect. Some cases with large $\alpha_\mathrm{ef}$ and $q$ even fall below the trajectory of the rigid-body rotation with $\tau_\mathrm{w,\odot}=6.6 \times 10^{30} \, \mathrm{erg} = \tau_\mathrm{w,\odot,M15}$ which is more than twice as large as $\tau_\mathrm{w,\odot,F19}$ \citep[black dashed line; ][]{Matt2015ApJ} to match the solar rotation along the evolution following the Skumanich law \citep[][see also Section \ref{sec:intro}]{Finley2018ApJ_Var1}. The solar rotation can be reproduced when $\alpha_\mathrm{ef} \gtrsim 0.40$ (redder lines) and $q \gtrsim 2$ (middle and right panels) even for the small $\tau_\mathrm{w,\odot}$. In other words, our model possibly reproduces the spin-down of both the Sun and solar-type stars with the in-situ measured solar wind torque in a self-consistent manner, when we assume one hypothesis: the Sun had the strong equator-fast DR until recently (see Section \ref{subsec:D_strong_solar_DR}).

Comparing the three panels of Fig. \ref{fig:SDE_1p0M}, we see that the scatter in $P_{\rm core}$ with respect to $\alpha_\mathrm{ef}$ is enhanced with larger value of $q$. This is because the effect of the LDR on the wind torque is boosted for larger $q$ (equation \ref{eq:re_X2}). Consequently, cases with large $\alpha_\mathrm{ef}$ experience more severe initial spin-down for larger $q$ (light blue arrows in Fig.~\ref{fig:schematic_trend}).

Readers may wonder if the above mentioned results are affected by the choice of the initial core angular velocity $\Omega_\mathrm{core,init}$. Fig.~\ref{fig:SDE_init} shows the spin evolution of the stars with $M= 1.0 \, \mathrm{M}_\odot$ for different $\Omega_\mathrm{core,init} = 2.7,10,27 \, \Omega_\mathrm{core,\odot}$. One can see that, even though the initial rotation velocities are different by ten times, the difference already converges within a factor of 1.5 at $t=1$ Gyr and to $\lesssim 10$ per cent after $t\gtrsim$ a few Gyr. It clarifies that the dependence of $\Omega_\mathrm{core,init}$ is so small that the discussion above is not significantly affected by the initial condition.

In order to determine how strong the solar equator-fast DR needs to be to reproduce the solar value, we search the possible pairs of $(q,\alpha_\mathrm{ef})$ that yield $\Omega_\mathrm{core} = \Omega_\mathrm{core,\odot}$ at $t=t_{\odot}$ for calibration. Fig.~\ref{fig:q_alphaefcal} shows the calibrated profile of $\alpha_\mathrm{ef,cal}$ in terms of $q$. The orange shaded region is for $2.7 \, \Omega_\mathrm{core,\odot} < \Omega_\mathrm{core,init} < 27 \, \Omega_\mathrm{core,\odot}$ with the solid line for $\Omega_\mathrm{core,init} = 10 \, \Omega_\mathrm{core,\odot}$. The derived value of $\alpha_\mathrm{ef,cal}(q)$ is within the constraint obtained from the highest observed value of $\Delta \Omega_\mathrm{core}/\Omega_\mathrm{eq}$ in solar analogues, KIC9025370 ($\Delta \Omega_\mathrm{core}/ \Omega_\mathrm{eq} \approx 0.35$-$0.69$; grey shaded region) and KIC8006161 ($\Delta \Omega_\mathrm{core}/\Omega_\mathrm{eq}\approx 0.14$-$0.29$; blue shaded region). This result suggests that our hypothesis of strong equator-fast DR of the young Sun does not contradict the observed data of the Sun and solar analogues.

\subsection{Mass dependence} \label{subsec:SDE_massdepend}

Fig.~\ref{fig:SDE_0p9M}, \ref{fig:SDE_1p1M} and  \ref{fig:SDE_1p2M} present the evolution of $\Omega_\mathrm{core}$ for  stars with $M= 0.9\, \mathrm{M_\odot}, 1.1 \, \mathrm{M_\odot}$ and $1.2 \, \mathrm{M_\odot}$. The distribution in $P_\mathrm{core}$ is broadened (narrowed) for $M= 0.9 \, \mathrm{M_\odot}$ ($M=1.1 \, \mathrm{M_\odot}$ and $1.2 \, \mathrm{M_\odot}$) than that for $M= 1.0 \, \mathrm{M_\odot}$. This trend can be understood as follows. A more massive star in the employed mass range has a shallower convection zone and hotter surface temperature, which gives shorter convective turn-over time ($\tau_\mathrm{cz}^\mathrm{CS}$) (see Fig.~\ref{fig:MESA_mass} in Appendix \ref{app:MESA}). In our prescription, the transition from equator-fast DR to pole-fast DR is set at the fixed Rossby number of Ro$_\mathrm{tran}$ described in Section \ref{sec:Model}. Since Ro $= P_\mathrm{core}/\tau_\mathrm{cz}^\mathrm{CS}$ (see equations \ref{eq:def_rossby} and \ref{eq:def_AM}), the rotation period at the transition is smaller for larger $M$. In other words, the transition from equator-fast DR to pole-fast DR occurs at earlier time for larger $M$. Therefore, a more massive star stays in the initial phase of the strong magnetic braking for shorter duration. This is the reason why the initial spin-down of the cases with $M=1.1 \, \mathrm{M_\odot}$ and $1.2 \, \mathrm{M_\odot}$ are not so significant as that of the lower-mass cases even though large $\alpha_\mathrm{ef}$ and $q$ are employed. Consequently, the vertical extent of the distribution is narrower for larger stellar mass as shown in Figure \ref{fig:SDE_1p1M}.  

Another characteristic feature regarding the mass dependence in our results is the concave down profile in spin-down of higher $M$ stars in the late phase; the stellar rotation declines sharply again after the spin-down is suppressed by the transition to the pole-fast DR. This is because of the effect of the stellar evolution. The stellar radius gradually expands even during the main sequence phase. The expansion timescale is shorter for higher $M$ because the stellar evolution proceeds more rapidly. As a result, the moment of inertia increases more rapidly in higher $M$ stars (Fig.~\ref{fig:MESA_mass} and Appendix \ref{app:MESA}). Therefore, the stellar rotation is decelerated dominantly by the change of $I$ (the second term of equation \ref{eq:AM_eq}) in the late phase even though the magnetic braking (the first term) is substantially suppressed then. 

The case with $M=1.2 \, \mathrm{M_\odot}$ exhibits a temporal rise of the rotation rate at $t\approx 4$ Gyr. A characteristic feature of this stellar-mass case is that the central part is occupied by a small convective region up to the mass radius of 0.05 M$_{\odot}$, unlike the lower $M$ cases. When the hydrogen is depleted in the central convection zone, the star contracts for a finite duration before nuclear burning is ignited in the shell outside the central region. Accordingly, the moment of inertia decreases during this period (see Appendix \ref{app:MESA} for detail), leading to spin-up (Fig.~\ref{fig:SDE_1p2M}). However, it should be noted that the spin-up is firstly induced in the core because the contraction occurs more severely in the inner region. Although our model is assuming that the spin-up of the core instantaneously affects the rotation on the surface (Section \ref{subsec:DR_model}), it is unclear whether this is actually realised because the efficiency of the transport of AM in the stellar interior is still under debate \citep{Aerts2019ARA&A}. Indeed, radial DR is detected in RGB stars, whose cores have experienced rapid contraction \citep[e.g.][]{Beck2012Nature}. If such radial DR is maintained, the temporal increase of the surface rotation seen in Fig. \ref{fig:SDE_1p2M} will be smeared out in realistic situations.

\subsection{Metallicity dependence} \label{subsec:SDE_metaldepend}

So far, we have been studying the spin-evolution of solar-metallicity stars. 
In order to investigate how the metallicity of stars affects the spin-evolution \citep[cf.][]{Amard2020ApJ}, we calculate the evolution of the stellar rotation with different metallicities, $Z=0.5,1.0$ and $2.0 \, \mathrm{Z}_\odot$ (see Section \ref{subsec:AM_evo} and Appendix \ref{app:MESA} for detail). Fig.~\ref{fig:SDE_metal} shows the results of the stars with $M=1.0 \, \mathrm{M_\odot}$ (top panel) , $M=1.1 \, \mathrm{M_\odot}$ (middle panel) and $M=1.2 \, \mathrm{M_\odot}$ (top panel) for each stellar abundance.

We can see that the metal-poor stars (blue lines) have rapider rotation than the stars with the solar abundance (black lines) throughout the evolution. The main reason for this metallicity dependence is that $\tau_\mathrm{cz}^\mathrm{CS}$ is smaller owing to the higher $T_{\rm eff}$ in the lower-metallicity condition (equation \ref{eq:timescale_overturn} and Fig.~\ref{fig:MESA_metal}), and thus, the wind torque, which is positively correlated  with $\tau_\mathrm{cz}^\mathrm{CS}$, (see equation \ref{eq:def_X2}) is reduced. While the modified $R(t)$ and $I(t)$ also affect the spin evolution (see equations \ref{eq:def_X1} and \ref{eq:AM_eq}, respectively), these factors do not significantly contribute. Inversely, the metal-rich (orange lines) stars follow an opposite trend, owing to large $\tau_\mathrm{cz}^\mathrm{CS}$. These tendencies are consistent with the results of \citet{Amard2020ApJ}, who extensively examined the impact of metallicity on the spin-evolution.

In addition, by comparing between the case with (solid lines in Fig~\ref{fig:SDE_metal}) and without (dotted lines in Fig~\ref{fig:SDE_metal}) LDR effect for each metallicity, we find that the LDR effect is less evident on stars with lower metallicity. It is caused by the earlier occurrence of the transition from equator-fast DR to pole-fast DR, owing to smaller $\tau_\mathrm{cz}^\mathrm{CS}$. This tendency is similar to that of higher-mass stars discussed in Section \ref{subsec:SDE_massdepend}.

\subsection{Observational test of the model}
\label{subsec:SDE_compareobs}

We have investigated the spin-evolution of solar-type stars with various masses and metallicities in Sections \ref{subsec:SDE_massdepend} and \ref{subsec:SDE_metaldepend}. Although the overall observational tendency can be explained by our results with the solar abundance, some stars are outside the range predicted by our calculations in spite of the wide ranges of the model parameters, $\alpha_\mathrm{ef}$ and $q$ (Figs. \ref{fig:SDE_1p0M}, \ref{fig:SDE_0p9M}, \ref{fig:SDE_1p1M} and \ref{fig:SDE_1p2M} in Section \ref{subsec:SDE_massdepend}). Even though, we can see the observational trend that the metal-poor (metal-rich) stars rotate fast (slowly) from Fig.~\ref{fig:SDE_metal} and that most of these outliers can be explained by this trend, following Section \ref{subsec:SDE_metaldepend}.

For example, KIC8394589, KIC9098294 and KIC10963065 in Fig.~\ref{fig:SDE_1p0M}, KIC8694723 in Fig.~\ref{fig:SDE_1p1M}, and KIC9965715 in Fig.~\ref{fig:SDE_1p2M} are rapidly rotating and positioned above the range of our model calculations using with the solar abundance. A characteristic point of these objects are that they are moderately metal-poor. Hence, this discrepancy might be solved by the ineffective spin-down due to the low metallicity (see the model with $Z= 0.5 \, \mathrm{Z}_\odot$ for $M = 1.0, 1.1$ and $1.2 \, \mathrm{M}_\odot$ in Fig.~\ref{fig:SDE_metal}). On the other hand, KIC9139151 in Fig.~\ref{fig:SDE_1p2M}, are located below the range of our model calculations with the solar metallicity. We can ascribe the effective spin-down of this star to the slightly smaller mass than $1.2 \, M_{\odot}$ and the slightly high metallicity.

However, although the low-metallicity model in Fig.~\ref{fig:SDE_metal} predicts the metal-poor star faces the transition from equator-fast DR to pole-fast DR earlier (see Section \ref{subsec:SDE_metaldepend}), three objects, KIC8694723, KIC9965715, and KIC10963065, of the five metal-poor stars have strong equator-fast DR even though they are near the terminal age of the main sequence. Moreover, KIC8179536 in Fig.~\ref{fig:SDE_1p2M} are rotating much more rapidly than our model calculation in spite of its solar-like metallicity. In order to solve these contradictions, further improvement of the model is required (see Section \ref{subsec:D_conv}).

Besides, there are old stars, KIC4914923 and KIC7680114 in Fig.~\ref{fig:SDE_1p1M}, locating near or after the terminal age of the main-sequence phase. Because these stars have slightly smaller masses than $1.1 \, M_{\odot}$ and slightly higher metallicities than the solar abundance, the main-sequence lifetime of these stars should be longer than that employed in the model calculations in Fig.~\ref{fig:SDE_1p1M} (see Appendix \ref{app:MESA}). Because of the sensitive dependencies of the lifetime on mass and metallicity in this mass range, the discrepancy can be partially explained by these slight deviations in the stellar masses and metallicities. Also, an adjustment of the detailed setting on the stellar evolution calculation (Appendix \ref{app:MESA}) could refine the model setup. When we elaborate theoretical models, old stars like these objects should be taken as important benchmarks (Section \ref{subsec:D_strong_solar_DR}).

\section{Discussion} \label{sec:Discussion}

\subsection{Uncertainty in Model Parameters} \label{subsec:D_Parameter}

We have examined the effect of LDR on the evolution of the rotation of solar-type stars. Since we employed simplified treatments for our analytic model (Section \ref{sec:Model}), there are several points that need to be addressed for future improvement.

We constructed the model for the time-evolution of LDR as a function of the Rossby number (equations~\eqref{eq:Omega_Rossby_trend} and~\eqref{eq:Omega_Rossby_formula} in Section \ref{subsec:DR_model}), which is defined by the empirical convective turn-over time $\tau^\mathrm{CS}_\mathrm{cz}$ \citep[equation \ref{eq:timescale_overturn};][]{Cranmer&Saar2011ApJ}. A caveat here is that $\tau^\mathrm{CS}_\mathrm{cz}$ is still uncertain. For example, \citet{See2021ApJ} compared several descriptions of $\tau^\mathrm{CS}_\mathrm{cz}$ and reported that the calculation with recent low-mass stellar models by \citet{Amard2019A&A} gives systematically longer $\tau^\mathrm{CS}_\mathrm{cz}$, and therefore smaller Ro, than \citet{Cranmer&Saar2011ApJ} by a factor $\lesssim 2$ for stars with $\sim 1 \, \mathrm{M_\odot}$. Although this does change the absolute value of Ro, we do not suppose that our model is substantially affected because the parameters, Ro$_{\rm tran}$ and Ro$_{\rm crit}$, in equations (\ref{eq:Omega_Rossby_trend}) and (\ref{eq:Omega_Rossby_formula})  are normalised by the solar value of Ro. 

On the other hand, we have set $\mathrm{Ro_{crit}}$ and $\mathrm{Ro_{tran}}$ to the fixed values for simplicity. 
While these values are chosen to explain the current observational properties as discussed in Section \ref{subsec:DR_model}, they may have to be modified when the observations are updated in future. 
However, it is supposed that the effect of a change in Ro$_{\rm crit}$ appears only in the early phase and the subsequent evolution is not significantly affected, which is similar to the dependence on the initial rotation shown in Fig. \ref{fig:SDE_init}. In contrast, a change in Ro$_{\rm tran}$ may substantially modify the evolution of $\Omega$, in combination with the parameters of equator-fast and pole-fast DR, $\alpha_{\rm ef}$ and $\alpha_{\rm pf}$. While we surveyed the evolutionary tracks with a range of $\alpha_{\rm ef}$, we fixed $\alpha_\mathrm{pf} = -3 \Delta \Omega_\mathrm{core,\odot} / \Omega_\mathrm{core,\odot}$, referring to the simulation data by \citet{Brun2022ApJ}. Here, we should note that it is not obvious whether a negative $\alpha_{\rm pf}$ is actually realised because stars with pole-fast DR have not been detected so far. However, although the precise determination of $\alpha_\mathrm{pf}$ is required to pin down the detailed late-phase evolutionary tracks in the $t-\Omega$ diagrams (Figures \ref{fig:SDE_1p0M}, \ref{fig:SDE_0p9M}, \& \ref{fig:SDE_1p1M}), we would like to emphasise that the trend of the weakened magnetic braking is generally maintained, provided that $\alpha_{\rm pf} \ll \alpha_{\rm ef}$, even when $\alpha_{\rm pf}\gtrsim 0$.

Our treatment of the wind torque is also a simplified one (Section \ref{subsec:Wind_model}). The parameter $q$, which determines the characteristic co-latitude of magnetic braking (Figure \ref{fig:q_thetaeff}), was assumed to be independent of time. In principle, $q$ should be calculated from the detailed structure of magnetised stellar winds \citep[e.g.][]{Ireland2022ApJ,FinleyBrun2023arXiv}. Hence, in general, $q$ is supposed to vary with time as the stellar magnetic configuration changes with stellar evolution. Since a decrease in $q$ also reduces the wind torque, this should be investigated in future studies. We also assumed the simple power-law dependence on stellar mass and radius with the same indices of $u$ and $v$ as in \citet[][]{Matt2015ApJ} (equation \ref{eq:def_X1}). While these indices are calibrated via the stellar observations in comparison to their model without LDR, they may have to be modified when the effect of LDR is considered.

\subsection{Strong equator-fast DR of the young Sun} \label{subsec:D_strong_solar_DR}

A critical hypothesis in our study is that the Sun exhibited the strong equator-fast DR until recently and is now in the transition phase. The Sun has lost its angular momentum quite efficiently in the past and is settled down to the current rotation state that is consistent with the small value of the in-situ measured solar-wind torque \citep{Finley2019ApJ_Direct,Finley2020ApJ}. This hypothesis is motivated by the observed solar-like stars with strong equator-fast DR \citep[][see also Figure \ref{fig:DRmodel}]{Benomar2018Sci}. The observed trend of magnetic activity suggesting that the Sun is under the transition also does not contradict our scenario \citep{Metcalfe&vanSaders2017SoPh}. 

However, one thing we should keep in mind is that the strong equator-fast DR detected in \citet{Benomar2018Sci} is not achieved in numerical simulations so far, although the current solar DR is reproduced without artificial settings quite recently \citep{Hotta2021NatAs, Hotta2022ApJ}.

This hypothesis is supposed to directly affect the solar and stellar magnetism because strong DR generally induces more active dynamo (see references in Section \ref{sec:intro}). Then, it indicates that the Sun had been magnetically active until recently. The propriety of the strong magnetic activity of the recent Sun may be tested in future by cosmochemical measurements of magnetised meteorites via solar winds \citep[e.g.,][]{Anand2022MNRAS,Obase2023Icar} and constraints given by the atmospheric escape from solar-system planets \citep[e.g.,][]{Terada2009AsBio,Cassata2022E&PSL}.

Another issue to be addressed is the variety of the evolution of the stellar rotation. We have been treating the `solar wind torque' as the referenced normalisation for the evolution of solar-type stars at the solar age. However, if the spin evolution of the Sun is deviated from the average trend of the bulk of solar-type stars, then the solar-wind torque should not be taken as the standard reference. Instead, it is required to construct a theoretical model that considers a finite distribution in the normalised wind torque in order to account for the variety of the spin evolution.

\subsection{Additional effect 1: effect of shallow surface convection zone}
\label{subsec:D_conv}

As discussed in Section \ref{subsec:SDE_metaldepend}, the elemental abundances are an essential player in determining the evolution of stars and their rotation (see also Fig.~\ref{fig:MESA_metal} in Appendix\ref{app:MESA}). As already emphasised by \citet{Amard2020ApJ}, higher(lower)-metallicity stars suffer more (less) efficient spin-down. We obtained the consistent result with them, and additionally, pointed out that several observational data deviated from the average trend is possibly explained by this metallicity effect (see Sections \ref{subsec:SDE_metaldepend} and \ref{subsec:SDE_compareobs}). For example, as shown in Section \ref{subsec:SDE_compareobs}, the five metal-poor stars in Figs. \ref{fig:SDE_1p0M}, \ref{fig:SDE_1p1M}, and \ref{fig:SDE_1p2M} exhibit relatively fast rotation.

However, there is a remaining issue on the discrepancy of the LDR condition. Asteroseismic observations detect fairly strong equator-fast DR in at least three objects of these five metal-poor stars (Table \ref{tab:landscape}). This contradicts our result because our model calculations predict that the phase of the strong equator-fast DR is already over in these stars.  
A possible solution to this discrepancy is an additional metallicity effect not considered in our model, which we discuss below.

The stellar metallicity and mass are also key parameters in determining the depth of the surface convection zone of stars especially in the supra-solar mass range of $1.0 \, {\rm M}_{\odot}\lesssim M \lesssim 1.5 \, {\rm M}_{\odot}$ \citep{Richard2002ApJ,Amard2019A&A}. Because the convective depth is expected to impact dynamo activity, the magnetic properties of such stars are regarded to be substantially affected by stellar metallicity and mass. 
In particular, the surface convection zone of lower-metallicity and higher-mass stars in this stellar-mass range is quite shallow or may almost disappear \citep{Richard2002ApJ}. In this case, the magnetic activity would be considerably suppressed owing to the absence of efficient dynamo in the convection zone. Accordingly, the magnetic braking would be also weakened, which could be an additional reason why metal-poor (blue symbols in Fig. \ref{fig:SDE_metal}) or high-mass (bottom panel in Fig. \ref{fig:SDE_metal}) stars show rapid rotation. 
It should be noted that this scenario can be identified with the observational tendency that the stars with $T_\mathrm{eff} > 6250 \, \mathrm{K}$ has rapid rotation \citep[cf. Kraft break;][]{Kraft1967ApJ}. In the forthcoming paper, we tackle the effect of stellar metallicity on the evolution of stellar rotation in more detail. 

\subsection{Additional effect 2: Metallicity-dependent stellar winds}
\label{subsec:D_metal}

Another issue regarding metallicity is that the physical condition of stellar winds is influenced by metallicity through radiative cooling in the stellar atmosphere. Radiative loss is suppressed in lower-metallicity gas because of a smaller number of coolants \citep[e.g.][]{SutherlandDopita1993ApJS} so that stronger stellar winds emanate. \citet{Suzuki2018PASJ} investigated Alfv\'{e}n-wave-driven stellar winds in different-metallicity conditions and derived a scaling relation of $\dot{M} \propto Z^{-1/5}$.

Considering equation (\ref{eq:scaling_tau_w}), we can derive the relation between the wind torque and metallicity as $\tau_\mathrm{w} \propto Z^{-(1-2\ell)/5}$.
By substituting $\ell=0.20$-$0.25$ (see the text after equation \ref{eq:scaling_tau_w}), the range of metallicity, $Z=0.5$-$2.0 \ \mathrm{Z}_\odot$, considered in this paper can alter $\tau_\mathrm{w}$ by $\lesssim 10$ per cent. Although the modification is almost negligible, this mechanism is a counter process against the metallicity effects previously discussed in Sections \ref{subsec:SDE_metaldepend} and \ref{subsec:D_conv} because it predicts stronger $\tau_{\rm w}$ for lower $Z$. Thus, the metallicity dependence from radiative cooling could in part cancel the reduced magnetic braking in low-metal stars reported by \citet{Amard2020ApJ}.

\subsection{Comparison with previous works} 
\label{subsec:D_other_mechanism}

In this paper, we proposed the transition of the DR profile as the origin of the weakened magnetic braking in the late evolutionary phase. On the other hand, as introduced in Section \ref{sec:intro}, \citet{vanSaders2016Nature} attribute the weakened magnetic braking to a change of magnetic geometry from a dipole field to higher-order fields. 
A simple model calculation that turns off the wind torque in the late phase nicely demonstrates the trend of their observational data. 

Although these two mechanisms appear to be differently originated, they are probably closely related \citep{Masuda2022ApJ}. For example, the transition in the rotational distribution inevitably has a significant impact on the amplification of the magnetic field. Inversely, the change in the magnetic topology influences the rotational properties through the magnetic braking. The mutual interaction between the dynamo process and the transport of AM is essentially important to understand the weakened magnetic braking.

\subsection{Error and bias of observation}
\label{subsec:D_errbiasobs}

In order to test our model, we compared our model calculations to the observed rotational periods in Section \ref{sec:observation} (see also Appendix \ref{app:SampleSelection}). The stellar rotation period can be measured by three different methods: 1. the spectroscopic measurement of rotational broadening, 2. the brightness modulation from starspots, and 3. the asteroseismic analysis about rotational frequency splitting. These different methods generally give consistent rotational periods in most objects \citep{Benomar2015MNRAS,Hall2021NatAs}. In this paper, we are adopting the result obtained by the methods 2. and 3., the spot modulation method and the asteroseismic method, because we focus on the core rotation rate.

What is probed by the spot modulation method is the rotation rates at active latitudes where the starspots are located. If we take sunspots for example, they appear mostly in latitude of $\lesssim 30^{\circ}$ \citep[e.g.][]{Maunder1904MNRAS,Hathaway2015LRSP}. If we can extrapolate the solar case to other stars,  the co-latitude $\theta=\theta_\mathrm{core} \approx 63.4^{\circ}$, ($=26.6^{\circ}$ in latitude) giving $\Omega_\mathrm{core}=\Omega_\mathrm{env}$ in our LDR model, is a reasonable value for the active latitude of solar-type stars. In this case, the rotation rate obtained with the spot modulation corresponds to that of the core on average. However, it is not clear whether the characteristics of sunspots can be directly applied to starspots particularly for stars with strong LDR. In addition, even in the Sun, the appearance latitude of sunspots varies within one solar-activity cycle. If this is the case for other solar-type stars, spot-modulation-based rotation rate would deviate from the core rotation rate, depending on the timing of observations. Then, we adopt the rotation rate obtained from the spot modulation just for the stars whose DR is not detected distinctly in \citet{Benomar2018Sci}. Besides,  although the spot modulation method probably selects the statistically plausible solution of rotation rate, multiple solutions or a solution with a large dispersion are sometimes derived for the rotation rate because of the presence of multiple starspots, the time-evolution of a starspot, or the variation from strong LDR \citep[][]{McQuillan2013MNRAS}. To avoid this error as much as possible, we eliminate the inconsistent value when comparing between the values obtained from the spot modulation method and the asteroseismic method (see sample selection in Appendix \ref{app:SampleSelection}).

The asteroseismic method can grasp the information of the core rotation more directly with a higher accuracy. However, it still involves errors arising from the limitation of observations. Although the asteroseismic method generally needs the information of various modes to derive the precise DR profile with the inversion technique \citep[see e.g.][for detail]{Aerts2010}, for the stars except for the Sun, we can only utilise a few modes because the amplitudes of the high-degree modes are too small to be identified in a light curve; positive and negative amplitudes of these geometrically complicated modes are cancelled out when integrated over the stellar surface. Owing to that, the model of the DR profile should be simple to obtain the reliable solution of rotation rate. Based on this background, we adopted the LDR model discussed in Section \ref{subsec:DR_model}, whose free parameters are only two variables, $\Omega_\mathrm{core}$ and $\Delta \Omega_\mathrm{core}$ (see equation \ref{eq:DRprofile_model}), with the assumptions that 1. the core is rigidly rotating, 2. the terms higher than the second order with respect to $\cos \theta $ are negligibly small in the envelope, and 3. the temporal variation and  radial dependence of rotation rate in the envelope are negligible. The assumption 3. is reasonable at least in the Sun because the helioseismic results show that the temporal and radial variation in the solar rotation rate in the envelope varies by only few percent \citep{Eff-Darwich2013SoPh}. On the other hand, while the assumptions 1. and 2. are already discussed in Section \ref{subsec:DR_model}, these may have to be modified when more detailed observational data are available.

\section{Summary and Conclusion} \label{sec:Conclusion}
The aim of our work is to pursue a solution to the two remaining problems on the evolution of the rotation of solar-type stars: (1) the wind torque inferred from the stellar observations being twice as large as that obtained from the in-situ solar observation and (2) the magnetic braking weakened in the late main-sequence phase. To this end, we constructed a phenomenological model about the spin evolution of solar-type stars considering the LDR effect.

We first developed a simple analytic model for the LDR profile that mimics the recent results of observations and numerical simulation. In particular, we implemented the two key characteristic trends: the rapid equator-fast DR in the early phase and the transition from equator-fast DR to pole-fast DR. Using this analytic description, we parameterised the strength of the magnetic braking in a way that it depends primarily on the magnetised stellar winds emanating from the low-latitude region.

We calculated the long-time evolution of the stellar rotation
with the LDR effect. In the early stage with the equator-fast DR, the stellar AM is more effectively carried off by the magnetised stellar winds than in the case without LDR. Eventually, the magnetic braking is suppressed as the rotational condition shifts from the equator-fast DR to the pole-fast DR.

In order to compare our model calculation to the observational data with different stellar masses and metallicities, we conducted the calculation with the stellar mass of $0.9, 1.0, 1.1$, and $1.2 \, \mathrm{M}_{\odot}$ and the stellar metallicity of $Z = 0.5, 1.0, $ and $2.0 \, \mathrm{Z}_{\odot}$. We found that the LDR effect is less evident for the star with higher stellar mass and lower metallicity because of the earlier termination of the strong equator-DR phase. Then, the current solar rotation and the average tendency of the rotation of solar-type stars can be naturally explained by the single value of the solar-wind torque directly measured in-situ. 
The observed variety of stellar rotation can be explained by the finite distribution of mass, metallicity, and LDR condition within the framework of our model. For example, our model presented the consistent results with the observational tendency that solar-type stars with lower (higher) metallicty have been observed to maintain rapider (slower) rotation.

These results suggest that the LDR effect can be a key to understanding the evolution of stellar rotation.

\section*{Acknowledgements}
We thank Allan Sacha Brun and Antoine Strugarek for providing the numerical data and insightful
comments for this study. T.T. is supported by IGPEES, WINGS Program, the University of Tokyo. T.K.S. is supported in part by Grants-in-Aid for Scientific Research from the MEXT/JSPS of Japan, 17H01105, 21H00033 and 22H01263 and by Program for Promoting Research on the Supercomputer Fugaku by the RIKEN Centre for Computational Science (Toward a unified view of the universe: from large-scale structures to planets, grant 20351188 -- PI J. Makino) from the MEXT of Japan.
M.S. is supported by the JSPS KAKENHI Grant Number JP22K14077.


\section*{Data availability}
The data underlying this article will be shared on reasonable request to the corresponding author.



\bibliographystyle{mnras}
\bibliography{example} 




\appendix

\section{Sample Selection} \label{app:SampleSelection}

In this section, we explain the selection of the observed data in this paper. Our observational samples are composed of Kepler Asteroseismic Science
Consortium (KASC) targets \footnote{http://kasoc.phys.au.dk/} and the well-studied solar-type stars: 16 Cyg A \& B. The basic stellar parameters, mass, age, surface gravity, effective temperature, and metallicity (Table \ref{tab:landscape}), based on asteroseismic modelling are adopted from \citet{Hall2021NatAs}. We extract the stars with $0.85$-$1.25 \, \mathrm{M}_\odot$ from their results because we focus on the solar-type stars.

The rotation rate of these stars is obtained from two different methods using the time variation of photometric light curves by starspots that rotate with a star \citep{Garcia2014AandA,Lu2022ApJ} and the rotational splitting of asteroseismic eigen-modes \citep{Benomar2018Sci,Bazot2019AandA}. In the former, which is referred to as the spot modulation method, the rotation period is calculated by using auto-correlation function \citep[ACF; e.g.][]{McQuillan2013MNRAS} and wavelet analysis \citep[WA; e.g.][]{Garcia2014AandA}. We adopt the observational data with WA in \citet{Garcia2014AandA} and \citet{Lu2022ApJ}; \citet{Garcia2014AandA} present the only data with WA although they compare those with ACF and WA; while \citet{Lu2022ApJ} present the both values from ACF and WA, we adopt the latter because these are more consistent with the values from the asteroseismic method (discussed in the later paragraphs).

The asteroseismic method enables us to obtain the LDR profile by information of multiple modes. The split frequencies of acoustic modes are characterised by Clebsch-Gordan $a$- coefficient $a_1, a_2$ and $a_3$; $a_1$ means the averaged rotation rate, $a_2$ indicates a measure of asphericity and $a_3$ corresponds to a measure of LDR \citep{Schou1994ApJ, Gizon2004SoPh, Gizon2016SciA}. \citet{Benomar2018Sci} and \citet{Bazot2019AandA} obtain the information of the LDR for KASC stars and 16 Cyg A \& B, respectively. We adopt the values of $a_1$ for all targets and $\Omega_\mathrm{core}$ and $\Omega_\mathrm{eq}$ for the stars with distinct LDR detected.

Through the selection above and excluding the sample with significantly large error, we compile the rotation and other stellar parameters about 31 samples by merging the results above.  We labelled the stars whose DR is detected distinctly by the asteroseismic method as `D' (Detected) and others as `N' (Non-detected). The obtained values are listed in Table \ref{tab:landscape}. It should be noted that the core rotation rates treated in Section \ref{sec:Result} are presented by the asteroseismic values (spot-modulation-based values) for the stars labelled `D'(`N').

\begin{figure}
	\includegraphics[width=0.9\columnwidth]{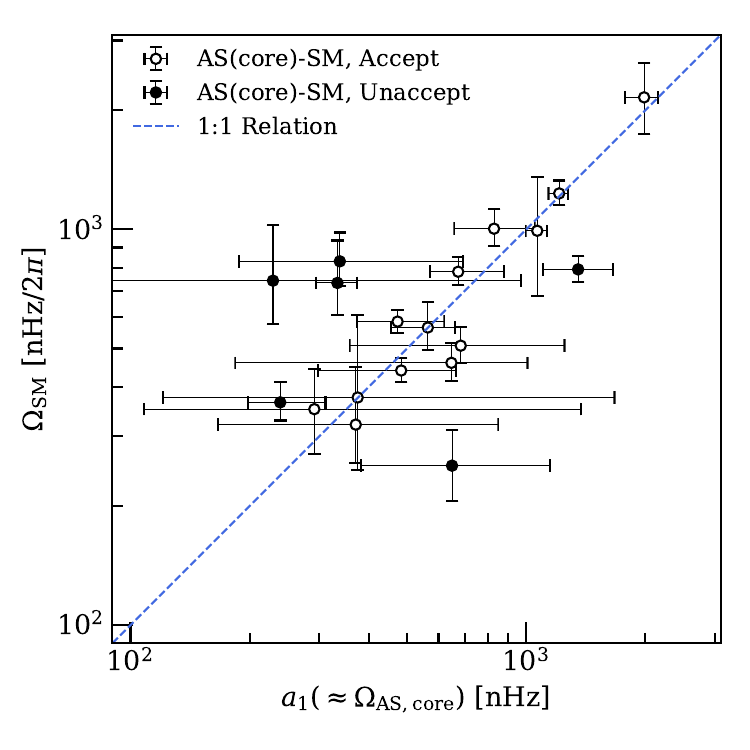}
    \caption{Direct comparison between $a$-coefficient $a_1$ (abscissa) and the angular velocity measured by the spot modulation method ($\Omega_\mathrm{SM}$; ordinate) about the stars whose LDR are not detected distinctly.  $a$-coefficient $a_1$ corresponds to the averaged rotation rate, which is nearly equal to the core rotation rate. The blue dashed line marks 1:1 ratio. The open (filled) circles with error-bars denote the acceptable (unacceptable) samples, which satisfy $a_1=\Omega_\mathrm{SM}$ including error-bars.
    }
    \label{fig:a1_vs_OmegaSM}
\end{figure}

However, as discussed in Section \ref{subsec:D_errbiasobs}, the rotation rates obtained by the spot modulation method ($\Omega_{\rm SM}$) may include the systematic error, depending on the latitudinal location of star spots. Then, we compare $\Omega_{\rm SM}$ with the core rotation rates obtained by the asteroseismic method ($\Omega_{\rm AS}$). Fig.~ \ref{fig:a1_vs_OmegaSM} shows the comparison between $a_1$ and $\Omega_{\rm SM}$ of the stars without clear detection of LDR. These samples are labelled `N', in Table \ref{tab:landscape}. Considering equation (\ref{eq:def_AM}), we can safely regard $a_1\approx \Omega_{\rm core}$ because $a_1$ corresponds to the averaged rotation rate . Fig.~ \ref{fig:a1_vs_OmegaSM} confirms that most of the samples (open circles with error-bars) have consistency between $a_1$ and $\Omega_{\rm SM}$, except for KIC3427720, KIC3656476, KIC6933899, KIC8150065, KIC8938364 and KIC10451113 (filled circles with error-bars). We remove these six stars from the discussion in Section \ref{sec:Result}.

\begin{figure}
	\includegraphics[width=0.9\columnwidth]{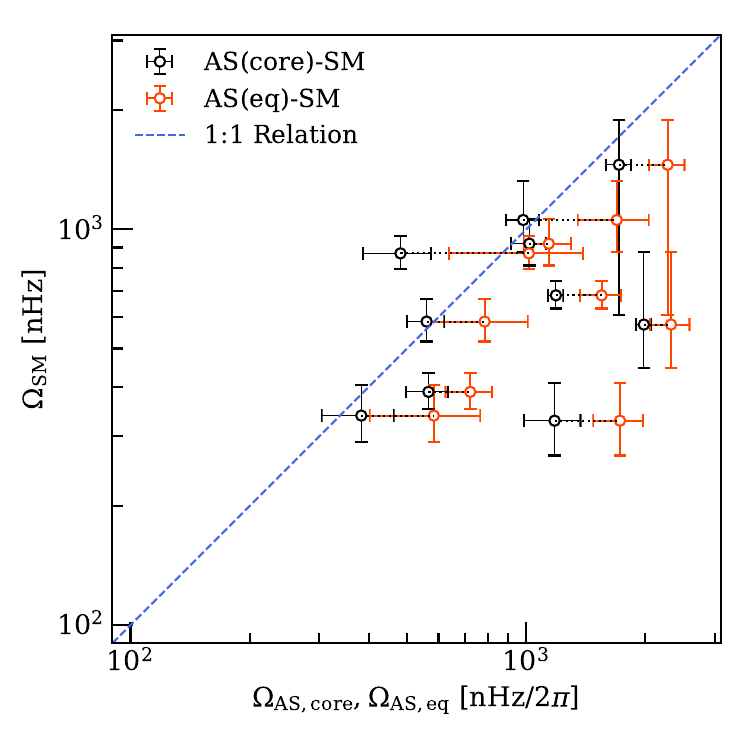}
    \caption{Direct comparison between the angular velocity measured by the asteroseismic method ($\Omega_\mathrm{AS}$; abscissa) and that measured by the spot modulation method ($\Omega_\mathrm{SM}$; ordinate) about the stars whose LDR are detected distinctly. The blue dashed line marks 1:1 ratio. Each star has two data points by the asteroseismic method that correspond to the rotation rate of the core, $\Omega_\mathrm{AS,core}$ (black circle with error-bars), and at the equator, $\Omega_\mathrm{AS,eq}$ (orange), respectively. $\Omega_\mathrm{AS,core}$ and $\Omega_\mathrm{AS,eq}$ of each star are connected by a black dotted line. }
    \label{fig:OmegaSM_vs_OmegaRS}
\end{figure}

Similarly, Fig.~\ref{fig:OmegaSM_vs_OmegaRS} shows the comparison between $\Omega_{\rm AS}$ and $\Omega_{\rm SM}$ of the stars with clear detection of equator-fast DR, which are labelled `D' in Table \ref{tab:landscape}. About half of the ten `D'-labelled samples satisfy $\Omega_{\rm SM}=\Omega_{\rm AS, core}$ within the range of the errors (black data points in Fig.~\ref{fig:OmegaSM_vs_OmegaRS}), while most of them give $\Omega_{\rm SM}<\Omega_{\rm AS, eq}$ (orange data points). These results support our assumption that the spot-modulation-based rotation rate gives a reasonable estimate for the core rotation rate (see discussion in Sections  \ref{sec:observation} and \ref{subsec:D_errbiasobs}).   
However, some stars show $\Omega_\mathrm{AS,core} > \Omega_\mathrm{SM}$, which is probably ascribed to the systematic error of the spot modulation method; we speculate that spots in polar regions, which are expected to rotate slowly, cause this discrepancy between $\Omega_\mathrm{AS,core}$ and  $\Omega_\mathrm{SM}$. Besides, there is a star which shows $\Omega_\mathrm{AS,core} < \Omega_\mathrm{SM}$. We note that this star, KIC9025370, has the strongest LDR in our samples, which may cause a scatter in the rotate rate based on the spot modulation method.

\begin{landscape}
 \begin{table}
  \caption{The list of the stellar values used in this paper.}
  \label{tab:landscape}
  \def\arraystretch{1.50}
  \setlength\tabcolsep{0.18cm}
  \begin{tabular}{ccccccccccccccccccc}
    \cline{1-18}
    
    & & \multicolumn{6}{c}{Fundamental Stellar Parameter} & &  \multicolumn{8}{c}{Rotation Rate} \\ 
    \cline{3-8} \cline{10-18} \\
    
    & & \multicolumn{6}{c}{Asteroseismic modelling} & & \multicolumn{2}{c}{Spot Modulation} & & \multicolumn{5}{c}{Asteroseismic Method} \\ 
    \cline{3-8} \cline{10-11} \cline{13-18} \\
    
    & $\mathrm{KIC}^{*1}$ & Mass &  {Age} & $\log g$  & $T_{\rm eff}$ & [Fe/H] & $\mathrm{Ref.}^{*5}$ & & $\Omega_{\rm SM} \,^{*2}$ & $\mathrm{Ref.}^{*6}$ & & $a_1$ & $\Omega_{\rm AS,core} \,^{*3}$ & $\Omega_{\rm AS,eq}$ & $\Delta \Omega_{\rm core} / \Omega_{\rm eq}$ & $\mathrm{Label}^{*4}$ &  $\mathrm{Ref.}^{*7}$ \\
    
    & & ($\mathrm{M}_\odot$) & (Gyr) & ($\mathrm{g \cdot cm^{-2}}$) & (${\rm K}$) & (dex) & & &  ($\mathrm{nHz}/2\pi$) &  & & ($\mathrm{nHz}/2\pi$) & ($\mathrm{nHz}/2\pi$) & ($\mathrm{nHz}/2\pi$) & & & \\
    
    \cline{1-18}
    
    & (3427720) & $1.11 ^{+0.02}_{-0.01}$ & $2.23^{+0.24}_{-0.24}$ & $4.386 ^{+0.006}_{-0.004}$ & 6045 $\pm$ 77 & -0.06 $\pm$ 0.1 &  [1] & & ($830^{+151}_{-111}$) & [2] & & $338^{+354}_{-150}$  & & & & N & [4] \\
    
    & (3656476) & $1.04 ^{+0.05}_{-0.04}$ & $8.37 ^{+1.72}_{-1.57}$ & $4.225 ^{+0.008}_{-0.010}$ & 5668 $\pm$ 77 & 0.25 $\pm$ 0.1 &  [1] & & ($365^{+46}_{-37}$) & [2] & & $239^{+72}_{-41}$ & & & & N & [4] \\
    
    & 3735871 & $1.09 ^{+0.04}_{-0.04}$ & $2.35 ^{+1.04}_{-0.85}$ & $4.396 ^{+0.007}_{-0.007}$  & 6107 $\pm$ 77 & -0.04 $\pm$ 0.1 &  [1] &  & $\mathbf{1004^{+121}_{-97}}$ & [2] & & $829^{+221}_{-171}$ & & & & N & [4] \\
    
    & 4914923 & $1.06 ^{+0.06}_{-0.05}$ & $7.57 ^{+1.66}_{-1.79}$ & $4.197 ^{+0.008}_{-0.010}$  & 5805 $\pm$ 77 & 0.08 $\pm$ 0.1 &  [1] &  & $\mathbf{565^{+90}_{-68}}$ & [2] &  & $564^{+97}_{-109}$ & & & & N & [4] \\
    
    & 5184732 & $1.15 ^{+0.04}_{-0.06}$ & $4.85 ^{+1.57}_{-0.88}$ & $4.255 ^{+0.010}_{-0.008}$  & 5846 $\pm$ 77 & 0.36 $\pm$ 0.1 &  [1] &  & $585^{+82}_{-64}$ & [2] &  & $552^{+60}_{-56}$ & $\mathbf{ 560^{+61}_{-61}}$ & $785^{+223}_{-223}$ & $0.29^{+0.17}_{-0.21}$ & D & [4]   \\

    & 5950854 & $0.97 ^{+0.03}_{-0.03}$ & $8.93 ^{+1.12}_{-1.15}$ & $4.238 ^{+0.007}_{-0.007}$  & 5853 $\pm$ 77 & -0.23 $\pm$ 0.1 &  [1] &  & $\mathbf{376^{+232}_{-129}}$ & [3] &  & $375^{+1296}_{-254}$ & & & & N & [4] \\
    
    & 6225718 & $1.16 ^{+0.03}_{-0.03}$ & $2.41 ^{+0.53}_{-0.43}$ & $4.319 ^{+0.006}_{-0.007}$  & 6313 $\pm$ 76 & -0.07 $\pm$ 0.1 &  [1] &  & ($329^{+80}_{-60}$) & [3] &  & $1141^{+226}_{-153}$ & $\mathbf{ 1179^{+192}_{-192}}$ & $1725^{+247}_{-247}$ & $0.32^{+0.7}_{-0.7}$ & D & [4] \\

    & 6603624 & $1.01 ^{+0.03}_{-0.02}$ & $7.82 ^{+0.94}_{-0.86}$ & $4.320 ^{+0.004}_{-0.006}$  & 5674 $\pm$ 77 & 0.28 $\pm$ 0.1 &  [1] &  & $\mathbf{351^{+93}_{-80}}$ & [3] &  & $291^{+1082}_{-183}$ & & & & N & [4] \\

    & (6933899) & $1.13 ^{+0.03}_{-0.03}$ & $6.34 ^{+0.72}_{-0.62}$ & $4.087 ^{+0.008}_{-0.007}$ & 5832 $\pm$ 77 & -0.01 $\pm$ 0.1 &  [1] &  & ($732^{+205}_{-125}$) & [3] &  & $333^{+41}_{-39}$ & & & & N & [4] \\
    
    & 7296438 & $1.08 ^{+0.06}_{-0.05}$ & $7.23 ^{+1.49}_{-1.77}$ & $4.320 ^{+0.009}_{-0.010}$ & 5775 $\pm$ 77 & 0.19 $\pm$ 0.1 &  [1] &  & $\mathbf{460^{+57}_{-46}}$ & [2] & & $647^{+360}_{-463}$ &  &  & & N & [4] \\
    
    & 7680114 & $1.06 ^{+0.04}_{-0.05}$ & $7.68^{+1.45}_{-1.28}$ & $4.172 ^{+0.008}_{-0.010}$ & 5811 $\pm$ 77 & 0.05 $\pm$ 0.1 &  [1] &  & $\mathbf{440^{+33}_{-29}}$ & [2] & & $483^{+182}_{-185}$ &  &  & & N & [4] \\
    
    & 8006161 & $0.98 ^{+0.03}_{-0.03}$ & $3.59 ^{+1.53}_{-1.45}$ & $4.494^{+0.007}_{-0.007}$ & 5488 $\pm$ 77 & 0.34 $\pm$ 0.1 &  [1] & & ($389^{+45}_{-37}$) & [2] & & $552^{+60}_{-56}$ & $\mathbf{ 566^{+69}_{-69}}$ & $722^{+97}_{-97}$ & $0.22^{+0.07}_{-0.08}$ & $\mathrm{D}^{\rm sa}$ & [4] \\

    & (8150065) & $1.19 ^{+0.04}_{-0.05}$ & $3.83 ^{+0.99}_{-0.67}$ & $4.220 ^{+0.008}_{-0.008}$ & 6173 $\pm$ 101 & -0.13 $\pm$ 0.15 &  [1] &  & ($253^{+59}_{-48}$) & [3] &  & $649^{+499}_{-266}$ & & & & N & [4] \\
    
    & 8179536 & $1.16 ^{+0.05}_{-0.06}$ & $3.54^{+1.04}_{-0.81}$ & $4.255 ^{+0.010}_{-0.010}$ & 6343 $\pm$ 77 & -0.03 $\pm$ 0.1 &  [1] &  & $\mathbf{2155^{+481}_{-417}}$ & [3] & & $1985^{+168}_{-212}$ &  &  & & N & [4] \\
    
    & 8379927 & $1.12 ^{+0.04}_{-0.04}$ & $1.99^{+0.85}_{-0.75}$ & $4.388^{+0.008}_{-0.007}$ & 6067 $\pm$ 120 & -0.1 $\pm$ 0.15 &  [1] & & ($681^{+59}_{-50}$) & [2] & & $1180^{+49}_{-57}$ & $\mathbf{ 1188^{+53}_{-53}}$ & $1550^{+185}_{-185}$ & $0.23^{+0.09}_{-0.09}$ & D & [4] \\

    & 8394589 & $1.04 ^{+0.04}_{-0.03}$ & $4.45^{+0.94}_{-0.83}$ & $4.322^{+0.008}_{-0.008}$ & 6143 $\pm$ 77 & -0.29 $\pm$ 0.1 &  [1] &  & $\mathbf{992^{+367}_{-315}}$ & [3] &  & $1066^{+61}_{-70}$ & & & & N & [4] \\

    & 8424992 & $0.92 ^{+0.04}_{-0.04}$ & $9.61^{+1.92}_{-1.74}$ & $4.359^{+0.007}_{-0.007}$ & 5719 $\pm$ 77 & -0.12 $\pm$ 0.1 &  [1] &  & $\mathbf{321^{+128}_{-65}}$ & [3] &  & $371^{+479}_{-204}$ & & & & N & [4] \\
    
    & 8694723 & $1.14 ^{+0.02}_{-0.02}$ & $4.69^{+0.48}_{-0.51}$ & $4.113^{+0.007}_{-0.009}$ & 6246 $\pm$ 77 & -0.42 $\pm$ 0.1 &  [1] & & $1456^{+429}_{-849}$ & [3] & & $1679^{+112}_{-122}$ & $\mathbf{ 1715^{+127}_{-127}}$ & $2276^{+234}_{-234}$ & $0.25^{+0.06}_{-0.07}$ & D & [4] \\

    & (8938364) & $0.99 ^{+0.01}_{-0.01}$ & $10.25^{+0.56}_{-0.65}$ & $4.173^{+0.007}_{-0.002}$ & 5677 $\pm$ 77 & -0.13 $\pm$ 0.1 &  [1] &  & ($742^{+281}_{-166}$) & [3] &  & $229^{+742}_{-142}$ & & & & N & [4] \\
    
    & 9025370 & $0.97^{+0.03}_{-0.03}$ & $6.55^{+1.26}_{-1.13}$ & $4.423^{+0.007}_{-0.004}$ & 5270 $\pm$ 180 & -0.12 $\pm$ 0.18 &  [1] & & ($870^{+94}_{-77}$) & [2] & & $469^{+87}_{-72}$ & $\mathbf{ 481^{+94}_{-94}}$ & $1015^{+378}_{-378}$ & $0.56^{+0.12}_{-0.22}$ & $\mathrm{D}^{\rm sa}$ & [4] \\
    
    & 9098294 & $0.97^{+0.02}_{-0.03}$ & $8.08^{+0.99}_{-0.73}$ & $4.308^{+0.006}_{-0.007}$ & 5852 $\pm$ 77 & -0.18 $\pm$ 0.1 &  [1] &  & $\mathbf{585^{+42}_{-37}}$ & [2] & & $473^{+148}_{-100}$ &  &  & & N & [4] \\
    
    & 9139151 & $1.18^{+0.04}_{-0.05}$ & $1.32^{+0.94}_{-0.75}$ & $4.382^{+0.008}_{-0.008}$ & 6302 $\pm$ 77 & 0.1 $\pm$ 0.1 &  [1] & & $1056^{+268}_{-178}$ & [2] & & $975^{+98}_{-97}$ & $\mathbf{ 982^{+94}_{-94}}$ & $1693^{+345}_{-345}$ & $0.42^{+0.11}_{-0.13}$ & D & [4] \\

     & 9410862 & $0.97^{+0.05}_{-0.04}$ & $6.93^{+1.49}_{-1.33}$  & $4.300^{+0.009}_{-0.008}$ & 6047 $\pm$ 77 & -0.31 $\pm$ 0.1 &  [1] & & $\mathbf{508^{+59}_{-48}}$ & [2] & & $683^{+566}_{-325}$ &  &  & & N & [4] \\
    
    & 9955598 & $0.90^{+0.04}_{-0.03}$ & $6.29^{+1.95}_{-1.84}$ & $4.497^{+0.007}_{-0.006}$ & 5457 $\pm$ 77 & 0.05 $\pm$ 0.1 &  [1] & & $338^{+67}_{-48}$ & [2] & & $383^{+80}_{-85}$ & $\mathbf{ 383^{+79}_{-79}}$ & $584^{+182}_{-182}$ & $0.31^{+0.23}_{-0.20}$ & D & [4] \\
    
    \cline{1-18}
    
  \end{tabular}
 \end{table}
\end{landscape}

\begin{landscape}
 \begin{table}
  \contcaption{}
  \def\arraystretch{1.50}
  \setlength\tabcolsep{0.18cm}
  \begin{tabular}{ccccccccccccccccccc}
    \cline{1-18}
    
    & & \multicolumn{6}{c}{Fundamental Stellar Parameter} & &  \multicolumn{8}{c}{Rotation Rate} \\ 
    \cline{3-8} \cline{10-18} \\
    
    & & \multicolumn{6}{c}{Asteroseismic modelling} & & \multicolumn{2}{c}{Spot Modulation} & & \multicolumn{5}{c}{Asteroseismic Method} \\ 
    \cline{3-8} \cline{10-11} \cline{13-18} \\
    
    & $\mathrm{KIC}^{*1}$ & Mass &  {Age} & $\log g$  & $T_{\rm eff}$ & [Fe/H] & $\mathrm{Ref.}^{*5}$ & & $\Omega_{\rm SM} \,^{*2}$ & $\mathrm{Ref.}^{*6}$ & & $a_1$ & $\Omega_{\rm AS,core} \,^{*3}$ & $\Omega_{\rm AS,eq}$ & $\Delta \Omega_{\rm core} / \Omega_{\rm eq}$ & $\mathrm{Label}^{*4}$ &  $\mathrm{Ref.}^{*7}$ \\
    
    & & ($\mathrm{M}_\odot$) & (Gyr) & ($\mathrm{g \cdot cm^{-2}}$) & (${\rm K}$) & (dex) & & &  ($\mathrm{nHz}/2\pi$) &  & & ($\mathrm{nHz}/2\pi$) & ($\mathrm{nHz}/2\pi$) & ($\mathrm{nHz}/2\pi$) & &  \\
    
    \cline{1-18}
    
    & 9965715 & $1.21^{+0.04}_{-0.05}$ & $2.92^{+0.86}_{-0.75}$ & $4.272^{+0.008}_{-0.009}$ & 5860 $\pm$ 180 & -0.44 $\pm$ 0.18 &  [1] & & ($575^{+301}_{-127}$) & [3] &  & $1967^{+84}_{-89}$ & $\mathbf{ 1982^{+89}_{-89}}$ & $2321^{+264}_{-254}$ & $0.15^{+0.08}_{-0.10}$ & D & [4] \\
    
    & 10079226 & $1.12^{+0.02}_{-0.03}$ & $3.06^{+0.70}_{-0.65}$ & $4.366^{+0.006}_{-0.006}$ & 5949 $\pm$ 77 & 0.11 $\pm$ 0.1 &  [1] & & $\mathbf{782^{+71}_{-60}}$ & [2] & & $673^{+205}_{-101}$ &  &  & & N & [4] \\
    
    & (10454113) & $1.17^{+0.02}_{-0.03}$ & $2.89^{+0.56}_{-0.53}$ & $4.314^{+0.006}_{-0.006}$ & 6177 $\pm$ 77 & -0.07 $\pm$ 0.1 &  [1] & & ($792^{+64}_{-55}$) & [2] & & $1352^{+307}_{-248}$ & & & & N & [4] \\
    
    & 10963065 & $0.99^{+0.06}_{-0.06}$ & $7.15^{+1.92}_{-1.61}$ & $4.277^{+0.011}_{-0.011}$ & 6140 $\pm$ 77 & -0.19 $\pm$ 0.1 &  [1] & & $920^{+144}_{-110}$ & [2] & & $1011^{+95}_{-98}$ & $\mathbf{ 1018^{+103}_{-103}}$ & $1140^{+155}_{-155}$ & $0.11^{+0.08}_{-0.10}$ & $\mathrm{D}^{\rm sa}$ & [4] \\
    
    & 12009504 & $1.17^{+0.02}_{-0.04}$ & $3.97^{+0.57}_{-0.43}$ & $4.211^{+0.007}_{-0.006}$ & 6179 $\pm$ 77 & -0.08 $\pm$ 0.1 &  [1] & & $\mathbf{1233^{+96}_{-83}}$ & [2] & & $1211^{+64}_{-73}$ &  &  & & N & [4] \\
    
    & 16 Cyg A & $1.05^{+0.02}_{-0.02}$ & $6.67^{+0.81}_{-0.77}$ & $4.287^{+0.007}_{-0.007}$ & 5825 $\pm$ 50 & 0.1 $\pm$ 0.03 &  [1] & & & & & $464^{+43}_{-43}$ & $\mathbf{471^{+43}_{-43}}$ & $535^{+76}_{-76}$ & $0.13^{+0.09}_{-0.10}$ & $\mathrm{D}^{\rm sa}$ & [5] \\
    
    & 16 Cyg B & $0.99^{+0.02}_{-0.02}$ & $7.39^{+0.89}_{-0.91}$ & $4.353^{+0.007}_{-0.006}$ & 5750 $\pm$ 50 & 0.05 $\pm$ 0.02 &  [1] & & & & & $531^{+91}_{-60}$ & $\mathbf{470^{+137}_{-104}}$ & $565^{+150}_{-129}$ & $0.15^{+0.11}_{-0.12}$  & $\mathrm{D}^{\rm sa}$ & [5] \\
    
    \cline{1-18}
    
    *1 & \multicolumn{17}{l}{The stars with parenthesis represent the stars whose $\Omega$ are only determined by the spot modulation method but the values are largely different from $a_1$ (see Fig.~ \ref{fig:a1_vs_OmegaSM}). These stars are eliminated from}\\
    & \multicolumn{17}{l}{the discussion in Section \ref{sec:Result}.} \\
    
    *2 & \multicolumn{17}{l}{The values with parenthesis are unacceptable due to large difference from $a_1$. The values written in bold style are adopted in the comparison of Section \ref{sec:Result}. } \\
    
    *3 & \multicolumn{17}{l}{The values written in bold style are adopted in the comparison of Section \ref{sec:Result}. } \\
    
    *4 & \multicolumn{17}{l}{The labels `D' (Detected) and `N' (Non-detcted) denote the stars whose LDR are detected and not detected distinctly, respectively. The suffix `sa' represent the solar-analogue stars }\\
    & \multicolumn{17}{l}{whose LDR are detected (see Fig.~\ref{fig:DRmodel}).} \\
    
    *5 & \multicolumn{17}{l}{Reference for stellar property: [1] \citet{Hall2021NatAs}.} \\
    *6 & \multicolumn{17}{l}{Reference for results from the spot modulation method: [2] \citet{Garcia2014AandA}; [3] \citet{Lu2022ApJ}.} \\
    *7 & \multicolumn{17}{l}{Reference for results from rotational frequency splitting: [4] \citet{Benomar2018Sci}; [5] \citet{Bazot2019AandA}. } \\
  \end{tabular}
 \end{table}
\end{landscape}


\section{Stellar evolution model calculated by MESA}
\label{app:MESA}

We determine the time-evolving stellar basic parameters with the \textsc{mesa} stellar evolution code. The adopted parameters and settings are listed in Table \ref{tab:mesa_setting}. We do not include any extra mixing processes such as diffusion and overshooting other than the normal convective mixing. Also, we ignore the mass loss by stellar winds because its contribution is generally negligible
\footnote{
When we use the empirical model of mass-loss rate \citep{Johnstone2015A&A}, the total mass loss integrated until 4.6 Gyr is less than 0.1 percent of the current solar mass \citep[][]{Fichtinger2017A&A}. However, if we enhanced mass-loss during the early phase was taken into account, the initial mass could be larger than the present value by 2 per cent \citep{Suzuki2013PASJ,Fichtinger2017A&A}. Nevertheless, we do not consider that the change in stellar mass is important in the spin-down because the second term of equation (\ref{eq:AM_eq}) is not as dominant as the first term during the early phase. Additionally, scatters in the evolution of rotation due to variations in the initial stellar properties rapidly converge to the average sequence, as illustrated in (Fig. \ref{fig:SDE_init}).}
in the main sequence phase and then we treat that the stellar mass is constant in time. The calculations of stellar evolution are conducted from the ZAMS (already defined in Section \ref{subsec:AM_evo}) to the terminal age of the main sequence when the central hydrogen abundance declines to $1 \times 10^{-5}$. While this setting reproduces well the physical properties of the present Sun, further scrutiny on each effect and parameter is needed for  more sophisticated models

\begin{figure}
	\includegraphics[width=\columnwidth]{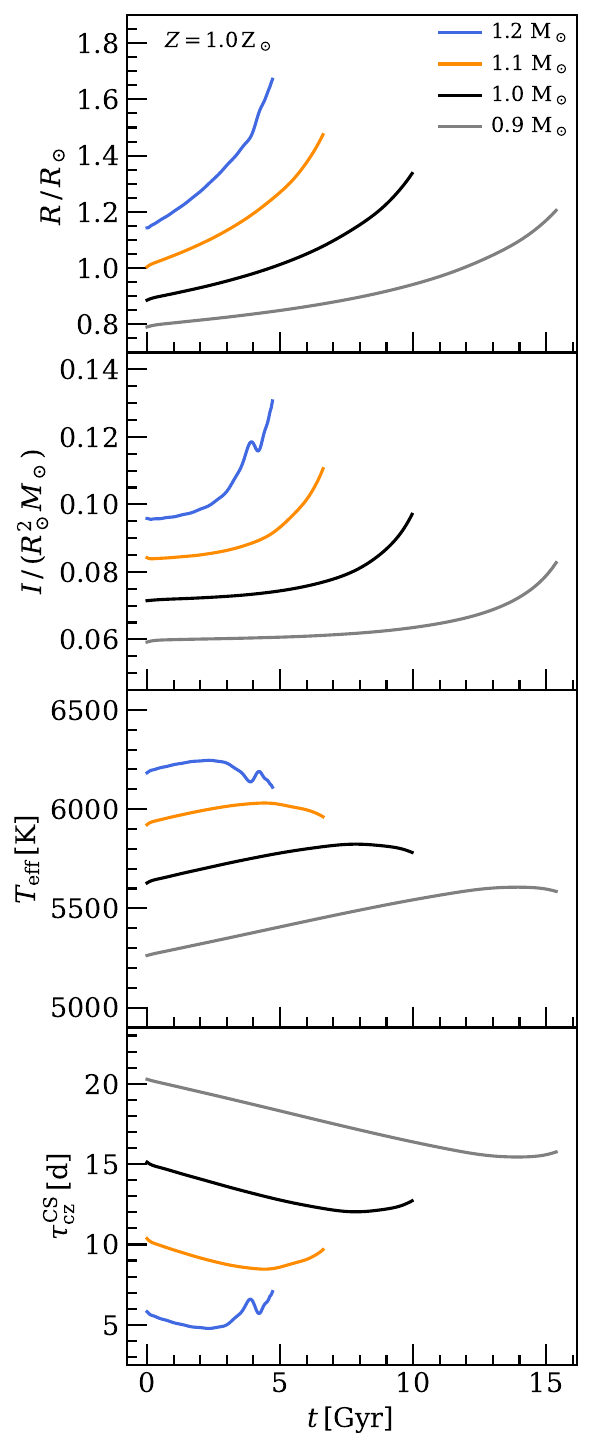}
    \caption{Time evolution of $R(t)$ (the first panel), $I(t)$(the second panel), $T_\mathrm{eff}(t)$ (the third panel) , $\tau_\mathrm{cz}^\mathrm{CS}(t)$ (the fourth panel) for  $M= 0.9 \, \mathrm{M}_\odot$ (grey lines), $1.0 \, \mathrm{M}_\odot$ (black lines),$1.1 \, \mathrm{M}_\odot$ (orange lines) and $1.2 \, \mathrm{M}_\odot$ (blue lines), respectively of solar-metallicity \citep{GS1998SSRv} stars.
    }
    \label{fig:MESA_mass}
\end{figure}

Fig.~\ref{fig:MESA_mass} shows the evolution of the radius $R(t)$, moment of inertia $I(t)$, effective temperature $T_\mathrm{eff}$, and convective turn-over time  $\tau_\mathrm{cz}^\mathrm{CS}(T_\mathrm{eff}(t))$ for the solar-abundance \citep[][]{GS1998SSRv} stars with $M= 0.9 \, \mathrm{M}_\odot$ (grey lines), $1.0 \, \mathrm{M}_\odot$ (black lines), $1.1 \, \mathrm{M}_\odot$ (orange lines) and $1.2 \, \mathrm{M}_\odot$ (blue lines), respectively. Because $\tau_\mathrm{cz}^\mathrm{CS}(T_\mathrm{eff})$ (the fourth panel of Fig.~\ref{fig:MESA_mass}) has the inverse correlation with the effective temperature (see definition in equation \ref{eq:timescale_overturn}), we only discuss the trend about the top three profiles. We can readily see the well-studied mass dependence from the figure: $R$ is larger; $I$ is larger; $T_\mathrm{eff}$ is higher for higher $M$ stars \citep[see e.g.][Chapter 20]{Kippenhahn2012}.

For $M \le 1.1 \,\mathrm{M}_\odot$, $R(t)$ (the first panel of Fig.~\ref{fig:MESA_mass}) and $I(t)$ (the second panel) rise over time, because the outer envelope gradually expands in accordance with the shrinkage of the core by the increase of the mean molecular weight due to the nuclear-fusion reaction in the central region. $T_\mathrm{eff}(t)$ (the third panels of Fig.~\ref{fig:MESA_mass}) increases in the early evolutionary phase owing to the increase of the nuclear burning rate, but it turns to decrease gently in the later phase along with the gradual transition to the off-centred nuclear burning, which is recognised as the main-sequence turn-off in a HR diagram  \citep[see e.g.][Chapter 5]{salaris2005}. On the other hand, the case with $M=1.2\,\mathrm{M}_\odot$ exhibits different behaviour from the lower $M$ cases because of the presence of the central convective region. One can see that $I(t)$ and $T_\mathrm{eff}(t)$ go up and down temporarily at $t \simeq 4 \, \mathrm{Gyr}$. It occurs from the contraction of the inner part of the star after the hydrogen depletion in the central convective zone that occupies $\sim 0.05 M_\odot$ at that time \citep[see e.g.][Chapter 5]{salaris2005}. The contraction continues during the star does not have a significant nuclear energy source before hydrogen burning is switched on in a shell. This phase corresponds to the period of the decrease in $I$ and the increase in $T_{\rm eff}$ in Fig.~\ref{fig:MESA_mass}. After the shell burning is turned on, $I$ increases and $T_{\rm eff}$ decreases again as in the lower $M$ cases.

\begin{figure}
	\includegraphics[width=\columnwidth]{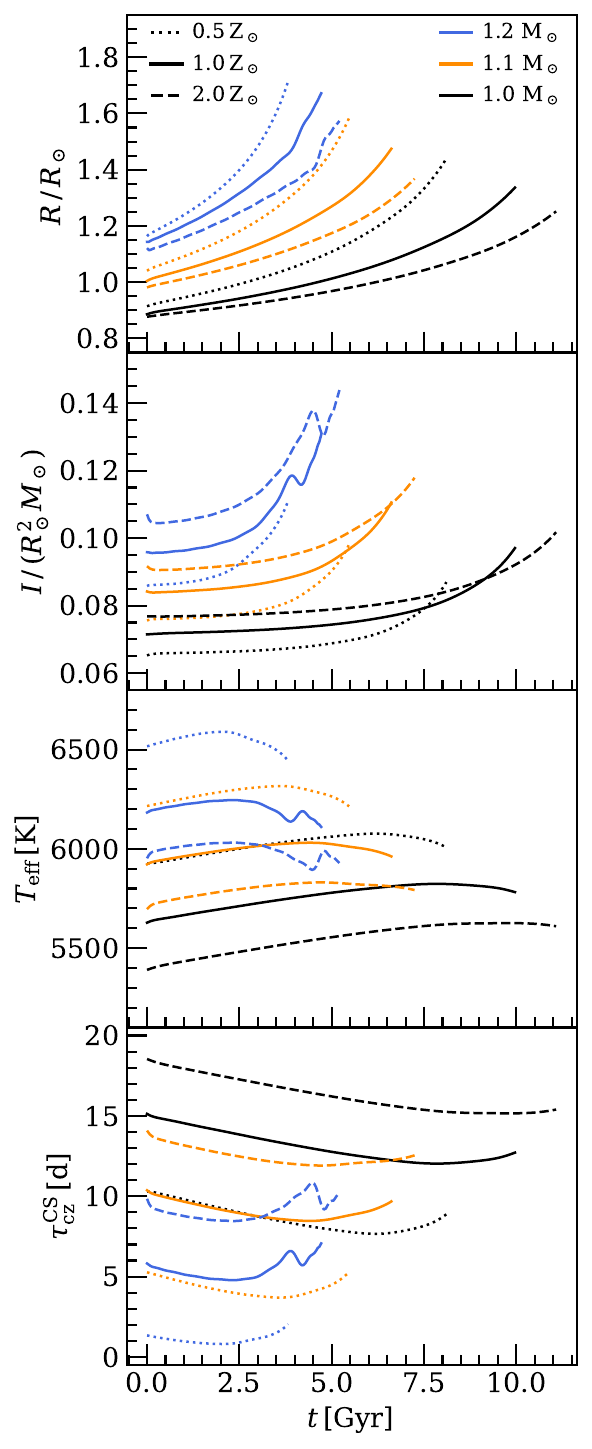}
    \caption{ Same as Fig. \ref{fig:MESA_mass} but for $M=1.0 \, \mathrm{M}_\odot$ (black lines), $M=1.1 \, \mathrm{M}_\odot$ (orange lines) and $1.2 \, \mathrm{M}_\odot$ (blue lines) with different metallicities, $Z=0.5 \, \mathrm{Z}_\odot$ (dotted lines), $1.0 \, \mathrm{Z}_\odot$ (solid lines), and $2.0 \, \mathrm{Z}_\odot$ (dashed lines), respectively (see Table \ref{tab:mesa_setting}). }
    \label{fig:MESA_metal}
\end{figure}

Next, we describe the dependence of the stellar properties on metallicity. Fig.~\ref{fig:MESA_metal} presents the results of $1.0 \, \mathrm{M_\odot}$, $1.1 \, \mathrm{M_\odot}$ and $1.2 \, \mathrm{M_\odot}$ stars with different heavy-element abundances, $Z=0.5 \, \mathrm{Z}_\odot$ (dotted lines), $Z=1.0 \, \mathrm{Z}_\odot$ (solid lines), and $Z=2.0 \, \mathrm{Z}_\odot$ (dashed lines), which are the same cases discussed in Section \ref{subsec:SDE_metaldepend}. Lower-metallicity stars give larger $R$ and higher $T_\mathrm{eff}$ because of more efficient nuclear burning. $I$ is smaller in spite of the larger $R$, because the density is more centrally concentrated. The star with lower (higher) metallicity has the shorter (longer) lifetime of the main-sequence phase because of the higher (lower) luminosity. These tendencies are consistent with the results obtained by \citet{Amard2020ApJ}.

\begin{table}
	\centering
	\caption{The setting of the \textsc{mesa}  stellar evolution code to calculate the time-evolving stellar basic parameters .
	}
	\label{tab:mesa_setting}
	\renewcommand{\arraystretch}{1.4}
	\begin{tabular}{lll} 
		\hline
		& Control name & Our setting \\ 
		\hline
		& Mass & $M = 0.9$, $1.0$, $1.1$ or $1.2 \, \mathrm{M}_{\odot}$ \\
		& Stellar abundance & $Z=1.0 \, \mathrm{Z}_\odot$; $Z=0.018$ and $Y=0.270^{*1}$   \\
		& & $Z=0.5 \, \mathrm{Z}_\odot$; $Z=0.009$ and $Y=0.252$   \\
		& & $Z=2.0 \, \mathrm{Z}_\odot$; $Z=0.036$ and $Y=0.306$   \\
		& Reaction net work & `pp\_and\_cno\_extras.net’ \\
		& Reaction rates & $\mathrm{JINA \ REACLIB}^{*2}$ \\
		& Atmosphere & `Eddington' \\
		& Mixing length parameter & $\alpha_\mathrm{MLT}=2.0$ \\
		& Diffusion & No \\
		& Overshooting & No \\
		\hline
		*1 & \multicolumn{2}{l}{\citet{GS1998SSRv}} \\
	    *2 & \multicolumn{2}{l}{\citet{Cyburt2010ApJS}}
	\end{tabular} 
\end{table}


\bsp	
\label{lastpage}
\end{document}